\documentclass[aps,twocolumn,superscriptaddress]{revtex4}
\usepackage{graphicx}

\newcommand{\bra}[1] {\left\langle #1 \right|}
\newcommand{\ket}[1] {\left| #1 \right\rangle}

\begin{document}

\title{Qubit-oscillator systems in the ultrastrong-coupling
regime and their potential for preparing nonclassical states}

\author{S. Ashhab}
\affiliation{Advanced Science Institute, The Institute of Physical
and Chemical Research (RIKEN), Wako-shi, Saitama 351-0198, Japan}
\affiliation{Physics Department, Michigan Center for Theoretical
Physics, The University of Michigan, Ann Arbor, Michigan
48109-1040, USA}

\author{Franco Nori}
\affiliation{Advanced Science Institute, The Institute of Physical
and Chemical Research (RIKEN), Wako-shi, Saitama 351-0198, Japan}
\affiliation{Physics Department, Michigan Center for Theoretical
Physics, The University of Michigan, Ann Arbor, Michigan
48109-1040, USA}

\date{\today}


\begin{abstract}
We consider a system composed of a two-level system (i.e.~a qubit)
and a harmonic oscillator in the ultrastrong-coupling regime,
where the coupling strength is comparable to the qubit and
oscillator energy scales. Special emphasis is placed on the
possibility of preparing nonclassical states in this system. These
nonclassical states include squeezed states, Schr\"odinger-cat
states and entangled states. We start by comparing the predictions
of a number of analytical methods that can be used to describe the
system under different assumptions, thus analyzing the properties
of the system in various parameter regimes. We then examine the
ground state of the system and analyze its nonclassical
properties. We finally discuss some questions related to the
possible experimental observation of the nonclassical states and
the effect of decoherence.
\end{abstract}


\maketitle

\section{Introduction}
\label{Sec:Introduction}

The two-level system (or qubit) and the harmonic oscillator are
the two most basic, and perhaps most often studied, components of
physical systems. The paradigm of a qubit coupled to a harmonic
oscillator has also been analyzed by numerous authors over the
past few decades \cite{JaynesCummings,GerryWalls}. Physical
systems that can be described by this model include natural atoms
coupled to optical or microwave cavities \cite{GerryWalls},
superconducting qubits coupled to superconducting resonators
\cite{Chiorescu,Wallraff,You}, quantum dots or Cooper-pair boxes
coupled to nanomechanical resonators
\cite{Hennessy,Leturcq,Armour}, electrons interacting with phonons
in a solid \cite{Holstein} and some models of chaotic systems
\cite{Graham}.

In the early work on cavity quantum electrodynamics (QED) in
atomic systems, the achievable atom-cavity coupling strengths were
smaller than the atomic and cavity decay rates, usually limiting
observations to only indirect signatures of the theoretically
predicted phenomena. Recently, the strong-coupling regime, where
the coupling strength is larger than the decay rates in the
system, has been achieved \cite{Kimble}. In addition to atomic
systems, the strong-coupling regime has been achieved in
superconducting circuit-QED systems \cite{Chiorescu,Wallraff}, and
superconducting-qubit-nanomechanical-resonator systems are
approaching this regime \cite{Leturcq}. In fact, superconducting
systems are suited for achieving the so-called
ultrastrong-coupling regime, where the qubit-oscillator coupling
strength is comparable to the qubit and oscillator energy scales
\cite{Devoret}. One can expect to find new phenomena in this
regime that are not present in the weak or moderately strong
coupling regimes. Indeed there have been a number of theoretical
studies on this system analyzing some of its rich static and
dynamical properties
\cite{Hines,Irish,Larson,Zueco,Meaney,Lizuain}.

One reason why superconducting systems are well suited for the
implementation of qubit-oscillator experiments is the flexibility
they allow in terms of designing the different system parameters.
For example, in the two earliest experiments on circuit QED,
Chiorescu {\it et al.}~\cite{Chiorescu} used a low-frequency
oscillator, while Wallraff {\it et al.}~\cite{Wallraff} realized a
resonant qubit-oscillator system. Sub-gigahertz qubits have also
been realized in recent experiments \cite{Oliver}, and there
should be no difficulty in fabricating high-frequency oscillators.
Therefore, all possible combinations of qubit and oscillator
frequencies are accessible, in principle. One advantage of
superconducting qubits over natural atoms is the additional
control associated with the tunability of essentially all the
qubit parameters \cite{Paauw}, as will be discussed in more detail
below. This tunability contrasts with the situation encountered
with natural atoms, where the atomic parameters are essentially
fixed by nature. This advantage can be seen clearly in the recent
experiments where Fock states and arbitrary oscillator states were
prepared in a superconducting qubit-oscillator system
\cite{Hofheinz,Houck}. We shall see, however, that the additional
controllability comes at the price of having to deal with
additional coupling channels to the environment, and this unwanted
coupling can increase the fragility of nonclassical states.

In this paper we present analytical arguments and numerical
calculations pertaining to the strongly coupled qubit-oscillator
system from the point of view of the potential for preparing
nonclassical states in this setup. These states include squeezed
states or superpositions of macroscopically distinct states
(i.e.~Schr\"odinger-cat state) in the oscillator, as well as
qubit-oscillator entangled states \cite{WeakCouplingProposals}. In
this study, we shall consider all the different combinations of
qubit and oscillator frequencies. We shall also analyze in some
detail the effect of the tunability in the qubit parameters on the
behaviour of the system.

The paper is organized as follows: In Sec.~\ref{Sec:Hamiltonian}
we introduce the Hamiltonian that we shall use throughout the
paper. In Sec.~\ref{Sec:AnalyticalMethods} we discuss various
analytical methods that can be used to study the system under
different assumptions, and we compare the predictions of these
methods. In Sec.~\ref{Sec:NumericalCalculations} we present
results of numerical calculations that demonstrate the properties
of the energy eigenstates of the system, including the
nonclassical properties of the ground state. In
Sec.~\ref{Sec:PreparationAndDetectionInSitu} we discuss the
possibility of preparing and detecting the three types of
nonclassical states of interest. In Sec.~\ref{Sec:Decoherence} we
discuss the effect of decoherence on the robustness of
nonclassical states. Section \ref{Sec:Conclusion} contains some
concluding remarks.

\section{Hamiltonian}
\label{Sec:Hamiltonian}

\begin{figure}[h]
\includegraphics[width=7.0cm]{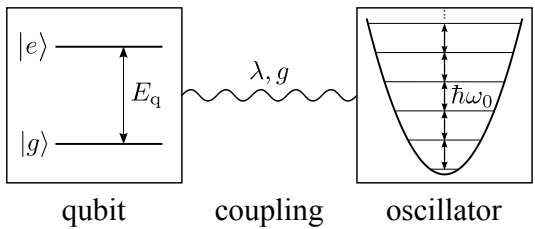}
\caption{Schematic diagram of the system under consideration. A
qubit with energy separation $E_{\rm q}$ between its ground and
excited states ($\ket{g}$ and $\ket{e}$) is coupled to a harmonic
oscillator with characteristic frequency $\omega_0$. The coupling
strength is given by $g$ or $\lambda$, depending on the language
used to describe the oscillator.} \label{Fig:SchematicDiagram}
\end{figure}

The system that we consider here is a qubit coupled to a harmonic
oscillator, as illustrated in Fig.~\ref{Fig:SchematicDiagram}.
Rather than worry about deriving the model from a microscopic
description of an electric circuit (see
e.g.~Refs.~\cite{Devoret,Bourassa,Peropadre}), we shall assume
that the description of the system as being composed of these two
physical components with a coupling term of the standard form is
an accurate description of the system.

The Hamiltonian of the system is given by:
\begin{equation}
\hat{H} = \hat{H}_{\rm q} + \hat{H}_{\rm ho} + \hat{H}_{\rm int},
\label{Eq:Basic_Hamiltonian}
\end{equation}
where
\begin{eqnarray}
\hat{H}_{\rm q} & = & -\frac{\Delta}{2} \hat{\sigma}_x -
\frac{\epsilon}{2} \hat{\sigma}_z \nonumber
\\
\hat{H}_{\rm ho} & = & \frac{\hat{p}^2}{2m} + \frac{1}{2}
m\omega_0^2 \hat{x}^2 \nonumber
\\
\hat{H}_{\rm int} & = & g \hat{x} \hat{\sigma}_z,
\label{Eq:Detailed_Hamiltonians}
\end{eqnarray}
$\hat{\sigma}_{x}$ and $\hat{\sigma}_{z}$ are the usual Pauli
matrices (with $\hat{\sigma}_z\ket{\uparrow}=\ket{\uparrow},
\hat{\sigma}_z\ket{\downarrow}=-\ket{\downarrow}$), and $\hat{x}$
and $\hat{p}$ are the position and momentum operators of the
harmonic oscillator. The parameters $\Delta$ and $\epsilon$ are
the gap and bias which characterize the qubit, $m$ is the
oscillator's effective mass, $\omega_0$ is the oscillator's
characteristic frequency, and $g$ is the qubit-oscillator coupling
strength. Note that in contrast to atomic cavity QED systems,
where $\epsilon=0$, this parameter is easily tunable in
present-day experiments using superconducting qubits. We shall
therefore treat $\epsilon$ as a tunable parameter (It is worth
noting here that most past studies on the ultrastrong-coupling
regime have focused on the case $\epsilon=0$; however, see
Ref.~\cite{Meaney}). For definiteness, we shall take $\Delta$ and
$g$ to be positive.

It is convenient for some calculations to express the oscillator
Hamiltonian using the creation ($\hat{a}^{\dagger}$) and
annihilation ($\hat{a}$) operators:
\begin{eqnarray}
\hat{a} & = & \hat{X} + i \hat{P} \nonumber
\\
\hat{a}^{\dagger} & = & \hat{X} - i \hat{P} \nonumber
\\
\hat{X} & = & \sqrt{\frac{m\omega_0}{2\hbar}} \hat{x} \nonumber
\\
\hat{P} & = & \frac{1}{\sqrt{2\hbar m\omega_0}} \hat{p} ,
\end{eqnarray}
which give
\begin{eqnarray}
\hat{H}_{\rm ho} & = & \hbar \omega_0 \hat{a}^{\dagger} \hat{a} +
\frac{1}{2} \hbar \omega_0 \nonumber
\\
\hat{H}_{\rm int} & = & \lambda \left( \hat{a} + \hat{a}^{\dagger}
\right) \hat{\sigma}_z \nonumber
\\
\lambda & = & g \sqrt{\frac{\hbar}{2m\omega_0}}.
\end{eqnarray}
The coupling strength can therefore be quantified either through
$g$ or $\lambda$.

We shall refer to the eigenstates of $\hat{H}_{\rm q}$ as the
qubit's ground and excited states, denoted by $\ket{g}$ and
$\ket{e}$, keeping in mind the caveat that this identification
becomes less meaningful for strong qubit-oscillator coupling. The
energies of these two states are $\pm E_{\rm q}/2$, where $E_{\rm
q}=\sqrt{\Delta^2+\epsilon^2}$. It is also useful to define an
angle $\theta$ that characterizes the relative size of the
$\hat{\sigma}_x$ and $\hat{\sigma}_z$ terms in the qubit
Hamiltonian: $\tan\theta=\epsilon/\Delta$. The eigenstates of
$\hat{H}_{\rm ho}$ are given by $\ket{n}$, where $n=0,1,2,...$,
with energies given by $n\hbar \omega_0$ (up to the ground state
energy $\hbar\omega_0/2$, which we ignore from now on). The
integer $n$ represents the number of excitations, to which we
shall refer as photons, in the oscillator.

\section{Comparison between different analytical methods}
\label{Sec:AnalyticalMethods}

In this section we describe some analytical methods that can be
used to determine the properties and behaviour of the system based
on different assumptions (which are valid in different parameter
regimes), and we compare the predictions of the different methods.

\subsection{Weak coupling}
\label{Sec:AnalyticalMethodsWeakCoupling}

The simplest limit is probably the weak-coupling limit
\cite{GerryWalls}, where $\lambda \ll \ E_{\rm q},\hbar\omega_0$.
Strictly speaking, one also needs to consider the number of
photons in the oscillator when determining whether the
weak-coupling condition is satisfied. However, since in this paper
we focus on a system that remains close to its ground state, we
assume a small number of photons in the oscillator. In the
weak-coupling limit, one can think of the qubit and oscillator as
being well-defined, separate physical systems that interact weakly
and can exchange excitations with one another
\cite{JaynesCummings}.

In the limit of small $\lambda$, one can treat $\hat{H}_{\rm int}$
as a small perturbation in the total Hamiltonian. The energy
eigenstates in the absence of this perturbation are given by
$\ket{n,g}=\ket{n}\otimes\ket{g}$ and
$\ket{n,e}=\ket{n}\otimes\ket{e}$, with energies
$n\hbar\omega_0\pm E_{\rm q}/2$ (recall that we ignore the
$\hbar\omega_0/2$ term in the oscillator's energy).

When there are no degeneracies in the non-interacting system
(i.e.~in the Hamiltonian given by $\hat{H}_{\rm q}+\hat{H}_{\rm
ho}$), the addition of the perturbation $\hat{H}_{\rm int}$ has
only a small effect on the behaviour of the system. This
perturbation only slightly modifies the energy levels and
eigenstates of the Hamiltonian.

When $E_{\rm q}\approx\hbar\omega_0$, the states $\ket{n,g}$ and
$\ket{n-1,e}$ are nearly degenerate (note that there is one such
pair of nearly degenerate states for each value of $n$), and the
perturbation term couples them. In particular, the relevant matrix
elements are given by $\bra{n,g}\hat{H}_{\rm
int}\ket{n-1,e}=\lambda \sqrt{n} \cos\theta$,
$\bra{n,g}\hat{H}_{\rm int}\ket{n,g}=\bra{n-1,e}\hat{H}_{\rm
int}\ket{n-1,e}=0$ and $\bra{n-1,e} (\hat{H}_{\rm q}+\hat{H}_{\rm
ho}) \ket{n-1,e} - \bra{n,g} (\hat{H}_{\rm q}+\hat{H}_{\rm ho})
\ket{n,g} = E_{\rm q}-\hbar\omega_0$. In other words the effective
Hamiltonian that one needs to consider is given by
\begin{equation}
\hat{H}_{\rm eff} = \left(\begin{array}{cc}
\delta/2 & \lambda\sqrt{n}\cos\theta \\
\lambda\sqrt{n}\cos\theta & -\delta/2
\end{array}
\right),
\end{equation}
where $\delta=E_{\rm q}-\hbar\omega_0$. Using this Hamiltonian,
one can analyze the behaviour of the system. In particular, when
$E_{\rm q}=\hbar\omega_0$, an excitation oscillates back and forth
between the qubit and oscillator with frequency $2 \lambda
\sqrt{n} \cos\theta$, which is commonly referred to as the Rabi
frequency.

Degeneracies also occur when $E_{\rm q}=k\hbar\omega_0$, with $k$
being any integer. In this case, one can go to higher orders in
perturbation theory and obtain analytic, though sometimes
cumbersome, expressions describing the properties and dynamics of
the system. We shall not go any further in analyzing this
situation here\cite{Shevchenko}.

Note that the same results as those explained above (for the case
$E_{\rm q}\approx\hbar\omega_0$) can be obtained by taking the
term $\hat{H}_{\rm int}$ in the Hamiltonian and replacing it by
its rotating-wave-approximation (RWA) form:
\begin{equation}
\hat{H}_{\rm int,RWA} = \lambda \cos\theta \left( \hat{a}
\hat{\sigma}_+ + \hat{a}^{\dagger} \hat{\sigma}_- \right),
\end{equation}
where $\hat{\sigma}_{\pm}$ are the qubit raising and lowering
operators ($\hat{\sigma}_+\ket{g}=\ket{e}$ etc.). This
approximation therefore ignores the so-called counter-rotating
terms in $\hat{H}_{\rm int}$, which are proportional to
$\hat{a}^{\dagger} \hat{\sigma}_+$ and $\hat{a} \hat{\sigma}_-$,
as well as a term proportional to
$(\hat{a}+\hat{a}^{\dagger})(\ket{e}\bra{e}-\ket{g}\bra{g})$ that
appears when $\epsilon\neq 0$. These terms would change the number
of excitations in the system, thus mixing states that have a large
energy separation (assuming $\lambda\ll E_{\rm q},\hbar\omega_0$),
and energy conservation suppresses such processes. The RWA
therefore approximates the original Hamiltonian by one where the
state $\ket{n,g}$ is coupled only to the state $\ket{n-1,e}$,
which would lead to exactly the same algebra and results mentioned
above. Some of the recent studies on ultrastrong coupling have
analyzed the effects of the counter-rotating terms on the system
dynamics \cite{Zueco,Lizuain}.

\subsection{High-frequency, adiabatically adjusting oscillator}
\label{Sec:AnalyticalMethodsAdiabaticOscillator}

The next limit that we consider is that where the oscillator's
characteristic frequency $\omega_0$ is large compared to the
qubit's energy splitting (i.e.~$\hbar\omega_0\gg E_{\rm q}$) and
also compared to the coupling strength
($\hbar\omega_0\gg\lambda$). In this case one can say that the
oscillator remains in its initial energy eigenstate (i.e.~ground
state, first excited state, etc.), and this state follows
adiabatically any changes in the qubit's state. This case was
analyzed theoretically in Refs.~\cite{Irish,Larson}.

The procedure for adiabatically eliminating the high-frequency
oscillator from the problem is straightforward. One starts by
noting that the qubit is coupled to the oscillator through the
operator $\hat{\sigma}_z$. As a result, one can think of the
oscillator as always monitoring the qubit observable $\sigma_z$
and adjusting to be in the instantaneous energy eigenstate that
corresponds to that value of $\sigma_z$ (Note here that if the
qubit is in a superposition of two different $\sigma_z$ states,
each part of the superposition -- with a well-defined value of
$\sigma_z$ -- will have the oscillator in the corresponding energy
eigenstate).

We therefore start by assuming that the qubit has a well-defined
value of $\sigma_z$, equal to $\pm 1$. The oscillator now feels
the effective Hamiltonian (calculated from $\hat{H}_{\rm ho}$ and
$\hat{H}_{\rm int}$):
\begin{equation}
\left. \hat{H}_{\rm ho, eff} \right|_{\sigma_z=\pm 1} = \hbar
\omega_0 \hat{a}^{\dagger} \hat{a} \pm \lambda \left( \hat{a} +
\hat{a}^{\dagger} \right).
\end{equation}
This Hamiltonian corresponds simply to the original oscillator
Hamiltonian with a constant force term applied to it. This force
term can be eliminated using the transformation
\begin{equation}
\hat{a}' = \hat{a} \pm \frac{\lambda}{\hbar\omega_0},
\end{equation}
which gives
\begin{equation}
\left. \hat{H}_{\rm ho, eff} \right|_{\sigma_z=\pm 1} = \hbar
\omega_0 \hat{a}'^{\dagger} \hat{a}' -
\frac{\lambda^2}{\hbar\omega_0}.
\end{equation}
The above steps can also be carried out in the language of the
operators $\hat{x}$ and $\hat{p}$:
\begin{eqnarray}
\left. \hat{H}_{\rm ho, eff} \right|_{\sigma_z=\pm 1} & = &
\frac{\hat{p}^2}{2m} + \frac{1}{2} m\omega_0^2 \hat{x}^2 \pm g
\hat{x} \nonumber
\\
x' & = & x \pm \frac{g}{m\omega_0^2} \nonumber
\\
p' & = & p \nonumber
\\
\left. \hat{H}_{\rm ho, eff} \right|_{\sigma_z=\pm 1} & = &
\frac{\hat{p'}^2}{2m} + \frac{1}{2} m\omega_0^2 \hat{x'}^2 -
\frac{g^2}{2m\omega_0^2}.
\end{eqnarray}

The energy levels of the oscillator are given by
$n\hbar\omega_0-\lambda^2/(\hbar\omega_0)$, independently of the
qubit's state. There will therefore not be a qubit-state-dependent
energy that we need to take into account when we turn to analyzing
the behaviour of the (slow) qubit. The oscillator's energy
eigenstates, however, are slightly dependent on the state of the
qubit. In particular,
\begin{equation}
\left\langle n_{\sigma_z=+1} | m_{\sigma_z=+1} \right\rangle =
\left\langle n_{\sigma_z=-1} | m_{\sigma_z=-1} \right\rangle =
\delta_{nm},
\end{equation}
and
\begin{widetext}
\begin{equation}
\left\langle n_{\sigma_z=+1} | m_{\sigma_z=-1} \right\rangle =
\left\{
\begin{array}{lll}
e^{-2\lambda^2/(\hbar\omega_0)^2}
\left(-\frac{2\lambda}{\hbar\omega_0}\right)^{m-n}
\sqrt{\frac{n!}{m!}} L_{n}^{m-n}\left[ \left(
\frac{2\lambda}{\hbar\omega_0} \right)^2 \right] & , & m\geq n \\
e^{-2\lambda^2/(\hbar\omega_0)^2}
\left(\frac{2\lambda}{\hbar\omega_0}\right)^{n-m}
\sqrt{\frac{m!}{n!}} L_{m}^{n-m}\left[ \left(
\frac{2\lambda}{\hbar\omega_0} \right)^2 \right] & , & m<n,
\end{array}
\right.
\end{equation}
\end{widetext}
where $\delta_{nm}$ is the Kronecker delta, and $L_i^j$ are the
associated Laguerre polynomials.

Having obtained the states of the high-frequency oscillator and
the properties of these states, one can now turn to the slow part
of the system, namely the qubit. We take any given value for the
index $n$, which specifies the oscillator's state, and we use it
to construct an effective qubit Hamiltonian for that value of $n$.
Since there are two qubit states for each value of $n$, the
effective Hamiltonian will be a $2 \times 2$ matrix operating in
the space defined by the states
$\left\{\ket{\widetilde{n,\uparrow}},\ket{\widetilde{n,\downarrow}}\right\}$
(We use the tildes in order to emphasize that the oscillator's
state is different from the state $\ket{n}$ of the free
oscillator). The four relevant matrix elements can be calculated
straightforwardly as
\begin{eqnarray}
\bra{\widetilde{n,\uparrow}} \hat{H}_{\rm q}
\ket{\widetilde{n,\uparrow}} & = & -
\bra{\widetilde{n,\downarrow}} \hat{H}_{\rm q}
\ket{\widetilde{n,\downarrow}} = -\frac{\epsilon}{2} \nonumber
\\
\bra{\widetilde{n,\uparrow}} \hat{H}_{\rm q}
\ket{\widetilde{n,\downarrow}} & = &
\bra{\widetilde{n,\downarrow}} \hat{H}_{\rm q}
\ket{\widetilde{n,\uparrow}} \nonumber
\\
& = & - \frac{\Delta}{2} e^{-2\lambda^2/(\hbar\omega_0)^2}
L_{n}^{0}\left[ \left( \frac{2\lambda}{\hbar\omega_0} \right)^2
\right].
\label{Eq:Renormalized_gap}
\end{eqnarray}
The qubit is therefore described by the effective Hamiltonian
\begin{equation}
\hat{H}_{\rm eff} = - \frac{1}{2} \left(\begin{array}{cc}
\epsilon & \tilde{\Delta} \\
\tilde{\Delta} & -\epsilon
\end{array}
\right),
\end{equation}
where $\tilde{\Delta}$ is given by Eq.~(\ref{Eq:Renormalized_gap})
and can be thought of as the renormalized value of the gap
$\Delta$.

The exponential and Laguerre-function factors are both slightly
smaller than one for small values of $\lambda/(\hbar\omega_0)$.
The qubit therefore experiences a small reduction in the coupling
(or `tunnelling') between the states $\ket{\uparrow}$ and
$\ket{\downarrow}$ in the weak-coupling limit
($\lambda\ll\hbar\omega_0$). This decrease in the renormalized
value of $\Delta$ can be understood in terms of the qubit having
to `pull' the oscillator with it as it tunnels between the states
$\ket{\uparrow}$ and $\ket{\downarrow}$, which would slow down the
tunneling process. Note that the renormalized gap depends on the
number of photons in the oscillator, which can lead to beating
dynamics and other interesting phenomena that occur when several
values of $n$ are involved \cite{Irish,Larson}.

If we keep increasing $\lambda/(\hbar\omega_0)$, without worrying
about satisfying the condition $\lambda\ll\hbar\omega_0$, we find
that the Laguerre polynomial and therefore the renormalized qubit
gap vanish at $n$ different $\lambda/(\hbar\omega_0)$ values (For
example, for $n=2$ there are two values of $\lambda$ for which the
renormalized gap vanishes). At these points, the states
$\ket{\widetilde{n,\uparrow}}$ and
$\ket{\widetilde{n,\downarrow}}$ are completely decoupled. Apart
from this feature, the renormalized gap decreases as a Gaussian
function with increasing $\lambda/(\hbar\omega_0)$. Note that
while increasing $\lambda$ there is no point that can be seen as a
`critical point' with a sudden change in behaviour. This situation
contrasts with what happens in the two calculations that we shall
discuss in the next two sections, and the meaning of the term
critical point will become clearer there.

One might expect that the adiabatically-adjusting-oscillator
approximation would break down when $\lambda$ is comparable to, or
larger than, $\hbar\omega_0$ (meaning that $\hbar\omega_0$ is not
the largest energy scale in the Hamiltonian). The arguments in the
previous paragraph might therefore seem of little significance. It
turns out, however, that the results discussed above hold, even
when $\lambda>\hbar\omega_0$. The reason why the approximation of
an adiabatically adjusting oscillator is still valid in this case
is that even though large changes in the oscillator's states can
occur when the qubit's state changes, these large changes involve
very slow processes that are governed by the renormalized gap. The
oscillator can therefore adjust adiabatically to these slow
processes. In other words, the condition that $\hbar\omega_0$ be
the largest energy scale in the Hamiltonian is a sufficient but
not necessary condition for the validity of the
adiabatically-adjusting-oscillator approximation.

\subsection{High-frequency, adiabatically adjusting qubit}
\label{Sec:AnalyticalMethodsAdiabaticQubit}

We now take the limit where $E_{\rm q}$ is much larger than both
$\hbar\omega_0$ and $\lambda$ (Some analysis of this case was
given in Ref.~\cite{Larson}). Similarly to what was done in
Sec.~\ref{Sec:AnalyticalMethodsAdiabaticOscillator}, we now say
that the qubit remains in the same energy eigenstate (ground or
excited state), and this state changes adiabatically following the
dynamics of the slow oscillator. We therefore start by finding the
energy eigenstates of the (fast-adjusting) qubit for a given state
of the (slow) oscillator. Since the interaction between the qubit
and the oscillator is mediated by the oscillator's position
operator $\hat{x}$, we start the calculation by assuming that $x$
has a well-defined value and treat the effective Hamiltonian
(obtained from $\hat{H}_{\rm q}$ and $\hat{H}_{\rm int}$):
\begin{equation}
\left. \hat{H}_{\rm q, eff} \right|_{x} = -\frac{\Delta}{2}
\hat{\sigma}_x - \frac{\epsilon}{2} \hat{\sigma}_z + g x
\hat{\sigma}_z.
\end{equation}
The eigenvalues and eigenstates of this Hamiltonian are given by:
\begin{eqnarray}
E_{\rm q, 1 | \it x} & = & - \frac{1}{2} \sqrt{\Delta^2 +
(\epsilon-2gx)^2} \nonumber
\\
\ket{g_x} & = & \cos\frac{\varphi}{2} \ket{\uparrow} +
\sin\frac{\varphi}{2} \ket{\downarrow} \nonumber
\\
E_{\rm q, 2 | \it x} & = & \frac{1}{2} \sqrt{\Delta^2 +
(\epsilon-2gx)^2} \nonumber
\\
\ket{e_x} & = & \sin\frac{\varphi}{2} \ket{\uparrow} -
\cos\frac{\varphi}{2} \ket{\downarrow} \nonumber
\\
\tan\varphi & = & \frac{\Delta}{\epsilon-2gx}.
\end{eqnarray}
We can now take these results and use them to analyze the
behaviour of the oscillator. We note here that since the variable
$x$ appears inside the square-root in the above expressions, the
operators $\hat{x}$ and $\hat{p}$ lead to a more transparent
analysis than the operators $\hat{a}$ and $\hat{a}^{\dagger}$. We
therefore use the operators $\hat{x}$ and $\hat{p}$ for the
remainder of this calculation.

Since the qubit's energy depends on the oscillator's position $x$,
the oscillator's effective potential now acquires a new
contribution (which depends on the qubit's state):
\begin{equation}
V_{\rm eff}(x) = \frac{1}{2} m\omega_0^2 x^2 \pm \frac{1}{2}
\sqrt{\Delta^2 + (\epsilon-2gx)^2}.
\label{Eq:effective_V_high_freq_qb}
\end{equation}
The plus sign corresponds to the qubit being in the excited state,
and the minus sign corresponds to the qubit being in the ground
state. In addition to the above effect of the qubit on the
oscillator, the qubit's state changes as the oscillator's position
changes, and the oscillator's kinetic-energy term will also be
modified in principle (this effect is similar to the
renormalization of $\Delta$ encountered in
Sec.~\ref{Sec:AnalyticalMethodsAdiabaticOscillator}). However, for
a sufficiently high-frequency qubit, changes in the qubit's state
will be small (see Appendix A), and consequently the change in the
kinetic-energy term can be neglected.

We now note that the effective potential in
Eq.~(\ref{Eq:effective_V_high_freq_qb}) is no longer a harmonic
potential. The second term describes one of the two branches of a
hyperbola, depending on the qubit's state. It will therefore not
be possible to derive general analytical results, and we have to
start considering some special cases.

In the limit $E_{\rm q}\gg g |x|$ for the relevant values of $x$,
the effective potential in Eq.~(\ref{Eq:effective_V_high_freq_qb})
can be approximated by:
\begin{eqnarray}
V_{\rm eff}(x) & \approx & \frac{1}{2} m\omega_0^2 x^2 \pm \left(
\frac{\sqrt{\Delta^2 + \epsilon^2}}{2} - \frac{\epsilon gx - g^2
x^2}{\sqrt{\Delta^2 + \epsilon^2}} \right) \nonumber
\\
& = & \frac{1}{2} m\tilde{\omega}_0^2 \left( x \mp \frac{\epsilon
g}{m\tilde{\omega}_0^2 E_{\rm q}} \right)^2 \pm \frac{E_{\rm
q}}{2},
\label{Eq:effective_V_high_freq_qb_Expansion}
\end{eqnarray}
where
\begin{equation}
\tilde{\omega}_0^2 = \omega_0^2 \pm 2 \frac{g^2}{m E_{\rm q}}.
\label{Eq:Renormalized_oscillator_frequency}
\end{equation}
The oscillator's effective potential is modified in two ways
depending on the qubit's state. Firstly, the location of the
minimum is shifted to the left or right by a distance proportional
to $\epsilon/E_{\rm q} = \sin\theta$ (This effect is absent when
the qubit is biased at the degeneracy point, where $\epsilon=0$).
Secondly, the oscillator's frequency is renormalized: According to
Eq.~(\ref{Eq:Renormalized_oscillator_frequency}), the oscillator's
effective frequency is increased for the qubit's excited state and
reduced for the qubit's ground state (This phenomenon is the basis
of the so-called quantum-capacitance and quantum-inductance
qubit-readout techniques \cite{Sillanpaa}).

An interesting result appears when one considers the case where
the qubit is in its ground state and
\begin{equation}
\frac{2g^2}{m\omega_0^2 E_{\rm q}} > 1, \hspace{0.5cm} {\rm i.e.}
\hspace{0.6cm} \frac{4\lambda^2}{\hbar\omega_0 E_{\rm q}} > 1.
\end{equation}
In this case the renormalized frequency becomes imaginary. This
result signals the presence of a critical point above which there
is an instability in the system. In particular, our expansion of
the square-root in
Eq.~(\ref{Eq:effective_V_high_freq_qb_Expansion}) is no longer
valid, the reason being that $x$ would increase indefinitely under
this approximation and the condition $E_{\rm q}\gg g|x|$ would be
violated.

The instability obtained above would raise questions about the
validity of the assumption of an adiabatically adjusting qubit
above the critical point. Nevertheless, we shall not worry about
this point now, and we continue the calculation. As a first step,
we note that $V_{\rm eff}(x)$ in
Eq.~(\ref{Eq:effective_V_high_freq_qb}) is well behaved at
$|x|\rightarrow\infty$. In particular,
\begin{equation}
V_{\rm eff}(x) = \frac{1}{2} m\omega_0^2 x^2 \pm \left| gx -
\frac{\epsilon}{2} \right| \hspace{0.4cm} {\rm when}
\hspace{0.4cm} |x| \gg \frac{\Delta}{g},\frac{|\epsilon|}{g}.
\end{equation}
We can therefore proceed with the calculation using the effective
potential given in Eq.~(\ref{Eq:effective_V_high_freq_qb}).

In order to make a few more statements about the case of strong
coupling (i.e.~above the critical point), it is useful to start
with the case $\epsilon=0$ and include a finite bias afterwards.
When $\epsilon=0$, the oscillator's effective potential takes the
form
\begin{equation}
V_{\rm eff}(x) = \left\{ \begin{array}{ll} \left( \frac{1}{2}
m\omega_0^2 \pm \frac{g^2}{\Delta} \right) x^2 \pm
\frac{\Delta}{2} & , x\ll \Delta/g
\\
\\
\frac{1}{2} m\omega_0^2 x^2 \pm |gx| & , x\gg \Delta/g
\end{array} \right.
\end{equation}
For the case with the qubit in its excited state (i.e.~when one
has the plus signs in the above expressions), the effective
potential is a slightly non-harmonic potential, and one can expect
the oscillator states to look more or less like the usual harmonic
oscillator states. For the case with the qubit in its ground
state, and when $2g^2/(m\omega_0^2\Delta)\ll 1$, one also has a
slightly non-harmonic potential. For the qubit's ground state and
$2g^2/(m\omega_0^2\Delta)> 1$ (which implies crossing the critical
point), the oscillator's effective potential is a double-well
potential. The locations of the minima can be obtained by setting
$dV_{\rm eff}/dx=0$ with $V_{\rm eff}$ given by
Eq.~(\ref{Eq:effective_V_high_freq_qb}): the minima are located at
$\pm x_0$, with
\begin{equation}
x_0 = \sqrt{\frac{g^2}{m^2\omega_0^4} - \frac{\Delta^2}{4g^2}}.
\label{Eq:Local_minimum_location}
\end{equation}
If one goes well beyond the critical point, the above expression
reduces to
\begin{equation}
x_0 \approx \frac{g}{m\omega_0^2},
\label{Eq:Local_minimum_location_Asymptotic}
\end{equation}
with minimum potential energy [measured relative to $V_{\rm
eff}(0)$]
\begin{equation}
V_{\rm min} = V_{\rm eff}(\pm x_0) - V_{\rm eff}(0) \approx -
\frac{g^2}{2m\omega_0^2},
\label{Eq:Energy_barrier}
\end{equation}
and curvature
\begin{equation}
\left. \frac{d^2V_{\rm eff}}{dx^2} \right|_{x=\pm x_0} \approx
m\omega_0^2.
\end{equation}
Note that this curvature is identical to that of the free
oscillator (i.e.~when $g=0$).

One can use the above expressions to estimate the energy
separation between the ground state and first excited state above
the critical point. These two states will be the symmetric and
antisymmetric superpositions of the ground states localized around
the two minima in the double-well potential. The distance between
the two minima is given by $2x_0$ from
Eq.~(\ref{Eq:Local_minimum_location_Asymptotic}), and the height
of the energy barrier separating the two minima is given by
$-V_{\rm min}$ from Eq.~(\ref{Eq:Energy_barrier}). Using the
Wentzel-Kramers-Brillouin (WKB) formula, one finds that the energy
separation between the two lowest states (and also within similar
pairs of higher energy levels) is exponential in the parameter
$\sqrt{-mV_{\rm min}} x_0/\hbar$, which is proportional to
$g^2/(m\omega_0^3\hbar)$, or alternatively
$\lambda^2/(\hbar\omega_0)^2$. This scaling is the same as the one
obtained in Sec.~\ref{Sec:AnalyticalMethodsAdiabaticOscillator}.

We now introduce $\epsilon$ to the problem. Far below the critical
point, the effect of $\epsilon$ can be obtained easily from
Eq.~(\ref{Eq:effective_V_high_freq_qb_Expansion}): the location of
the minimum in the single-well effective potential is slightly
shifted to the left or right. More care is required above the
critical point, where one has the double-well effective potential.
In this case, a finite value of $\epsilon$ breaks the symmetry
between the left and right wells, thus giving an energetic
preference for one of the two wells. In order to cause
localization in the energy eigenstates, $\epsilon$ has to be
larger than the energy separation within one of the energy-level
pairs discussed above, i.e.~$\epsilon$ needs to be larger than a
quantity that is exponentially small in $g^2/(m\omega_0^3\hbar)$.
Clearly, this localization happens at smaller values of $\epsilon$
as one goes deeper into the bistability region. This result means
that the superpositions involving both wells become increasingly
fragile with increasing coupling strength.

Finally we note that above the critical point, one finds that the
condition $E_{\rm q}\gg g|x|$ can no longer be satisfied for any
of the energy eigenstates. Therefore, one might expect that the
present approximation cannot be trusted. As we discussed in
Sec.~\ref{Sec:AnalyticalMethodsAdiabaticOscillator}, however, the
above results hold even when $\lambda>E_{\rm q}$. In that case the
energy eigenstates are either localized close to one of the local
minima or involve very slow tunneling between the two wells of the
effective double-well potential. The qubit can adjust
adiabatically to such slow tunneling processes. It is worth
mentioning here that when $\lambda$ is the largest energy scale in
the Hamiltonian, the approximations of this section and
Sec.~\ref{Sec:AnalyticalMethodsAdiabaticOscillator} are both
valid, and either one of the two approaches can be used to answer
any given question.

\subsection{Semiclassical calculation}
\label{Sec:AnalyticalMethodsSemiclassical}

\begin{figure}[h]
\includegraphics[width=8.5cm]{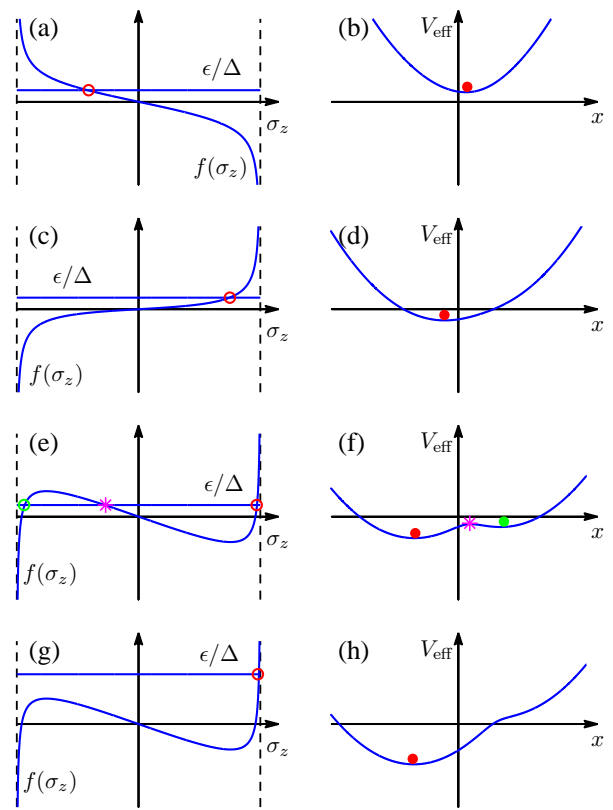}
\caption{(Color online) Graphical solution of
Eq.~(\ref{Eq:Semiclassical_solution}) (left) and the associated
effective potentials from
Sec.~\ref{Sec:AnalyticalMethodsAdiabaticQubit} (right) in the
different possible cases. The horizontal lines in the left panels
represent the first term in Eq.~(\ref{Eq:Semiclassical_solution}),
and the function $f(\sigma_z)$ is the second term in the equation.
The circles mark the stable solutions of
Eq.~(\ref{Eq:Semiclassical_solution}), and the dots in the right
panels mark the minima in the effective potential of
Sec.~\ref{Sec:AnalyticalMethodsAdiabaticQubit},
i.e.~Eq.~(\ref{Eq:effective_V_high_freq_qb}). The magenta stars in
panels (e) and (f) mark an unstable stationary point, i.e.~a local
maximum in the effective potential. Panels (a) and (b) correspond
to the plus signs in Eqs.~(\ref{Eq:effective_V_high_freq_qb}) and
(\ref{Eq:Semiclassical_solution}), and panels (c-h) correspond to
the minus signs. In panels (c) and (d), the coupling is below the
critical point, i.e.~$2g^2<m\omega_0^2E_{\rm q}$. In panels (e)
and (f), the coupling is above the critical point and $\epsilon$
is small. In panels (g) and (h), the coupling is above the
critical point and $\epsilon$ is large.}
\label{Fig:SemiClassicalGraphicalSolution}
\end{figure}

A semiclassical calculation can go as follows (alternative
semiclassical calculations can be found in
\cite{Graham,Hines,Larson}): The five different variables $x$,
$p$, $\sigma_x$, $\sigma_y$ and $\sigma_z$ are treated as
classical variables whose dynamics obeys the Hamiltonian in
Eqs.~(\ref{Eq:Basic_Hamiltonian}) and
(\ref{Eq:Detailed_Hamiltonians}), without the hats. These
variables obey the constraint
$C=\sigma_x^2+\sigma_y^2+\sigma_z^2=1$. One can therefore find the
ground state relatively easily by minimizing the Hamiltonian under
the above constraint. Minimizing the function $\tilde{H}=H-\mu C$,
with $\mu$ being a Lagrange multiplier, results in the set of
equations
\begin{eqnarray}
\frac{d\tilde{H}}{dx} & = & m \omega_0^2 x + g \sigma_z = 0
\nonumber
\\
\frac{d\tilde{H}}{dp} & = & \frac{p}{m} = 0 \nonumber
\\
\frac{d\tilde{H}}{d \sigma_x} & = & - \frac{\Delta}{2} - 2 \mu
\sigma_x = 0
\label{Eq:Semiclassical_equations}
\\
\frac{d\tilde{H}}{d \sigma_y} & = & - 2 \mu \sigma_y = 0 \nonumber
\\
\frac{d\tilde{H}}{d \sigma_z} & = & - \frac{\epsilon}{2} + g x - 2
\mu \sigma_z = 0, \nonumber
\end{eqnarray}
which are to be solved under the constraint
$\sigma_x^2+\sigma_y^2+\sigma_z^2=1$. The first four equations
lead to $x=-g\sigma_z/(m\omega_0^2)$, $p=0$,
$\mu=-\Delta/(4\sigma_x)$, and $\sigma_y=0$. The constraint gives
$\sigma_x=\pm\sqrt{1-\sigma_z^2}$. One is therefore left with the
equation
\begin{equation}
- \frac{\epsilon}{2} - \frac{g^2\sigma_z}{m\omega_0^2} \pm
\frac{\Delta \sigma_z}{2\sqrt{1-\sigma_z^2}} = 0,
\end{equation}
which can be re-expressed as
\begin{equation}
- \frac{\epsilon}{\Delta} - \left( \frac{2g^2}{m\omega_0^2\Delta}
\pm \frac{1}{\sqrt{1-\sigma_z^2}} \right) \sigma_z = 0.
\label{Eq:Semiclassical_solution}
\end{equation}
The above equation cannot be solved in closed form, in general.
However, one can make some general statements about the solution
(see Fig.~\ref{Fig:SemiClassicalGraphicalSolution}). For the plus
sign (Fig.~\ref{Fig:SemiClassicalGraphicalSolution}a), the second
term in Eq.~(\ref{Eq:Semiclassical_solution}) is a monotonically
decreasing function that approaches $+\infty$ when $\sigma_z
\rightarrow -1$ and approaches $-\infty$ when $\sigma_z
\rightarrow 1$. There is therefore one solution to the equation in
that case. It turns out that this solution does not correspond to
the ground state (This fact can be seen as simply a result of
comparing the energies of the different solutions). The ground
state is obtained when using the minus sign. In this case, there
are three possibilities: The first possibility is to have
$2g^2<m\omega_0^2\Delta$. In this case
(Fig.~\ref{Fig:SemiClassicalGraphicalSolution}c), we find that the
second term in Eq.~(\ref{Eq:Semiclassical_solution}) is a
monotonically increasing function that approaches $-\infty$ when
$\sigma_z \rightarrow -1$ and approaches $+\infty$ when $\sigma_z
\rightarrow 1$. There is therefore only one solution to
Eq.~(\ref{Eq:Semiclassical_solution}). The second and third
possibilities for the solutions of
Eq.~(\ref{Eq:Semiclassical_solution}) occur when
$2g^2>m\omega_0^2\Delta$. In this case
(Figs.~\ref{Fig:SemiClassicalGraphicalSolution}e and
\ref{Fig:SemiClassicalGraphicalSolution}g), the second term in
Eq.~(\ref{Eq:Semiclassical_solution}) develops a local maximum and
a local minimum between $\sigma_z=-1$ and $\sigma_z=1$. Depending
on the value of $\epsilon$, there can be either one or three
solutions. In particular, when $\epsilon=0$, the three solutions
are given by $\sigma_z=0$, which turns out to be an unstable
stationary point, and $\sigma_z=\pm
\sqrt{1-(m\omega_0^2\Delta/2g^2)^2}$, which are two degenerate
ground states (It is easy to verify that this result agrees with
Eq.~\ref{Eq:Local_minimum_location}).

One can intuitively understand the effect of having a finite value
of $\epsilon$ using the language of
Sec.~\ref{Sec:AnalyticalMethodsAdiabaticQubit}. For $\epsilon=0$,
one has an effective trapping potential for the variable $x$, and
this potential has the shape of a harmonic-oscillator-like
single-well potential when $2g^2<m\omega_0^2\Delta$ and a
double-well potential when $2g^2>m\omega_0^2\Delta$. This
situation explains the existence of one ground state when
$2g^2<m\omega_0^2\Delta$ and two degenerate ground states when
$2g^2>m\omega_0^2\Delta$. The effect of adding $\epsilon$ to the
problem is to create a tilt in the effective trapping potential; a
positive value of $\epsilon$ favours the negative-$x$ solution
(here we have in mind the ground-state solution). If the tilt is
weak, one has a global minimum in the deeper well and a local
minimum in the shallower well. If the tilt exceeds a certain
critical value, the shallow well is eliminated, and one recovers a
single-well potential.

We make a final note on the fact that we started the calculation
by raising a question related to the ground state but found
multiple solutions. The reason for this result is the fact that
Eq.~(\ref{Eq:Semiclassical_equations}) locates all stationary
points, and not only the ground state. The calculation therefore
identifies both the ground state and also high-energy stationary
points that are either dynamically unstable or dynamically stable
but can still relax to lower-energy states.

\subsection{Concluding remarks}

We conclude this section with some remarks on the conceptual ideas
and predictions of the different analytical methods. The standard
perturbation-theory procedure is well suited for the weak-coupling
limit. One can use it to systematically obtain accurate
approximations for the energy eigenstates and dynamics of the
system. The two approximations involving one subsystem, either the
qubit or the oscillator, adjusting adiabatically to the slow
dynamics of the other one are based on the conceptual picture of
the separation between different time scales in the problem. As
formulated above, they involve only one level of approximation, in
contrast to the order-by-order expansion involved in perturbation
theory. The time-scale-separation-based approximations can,
however, be constructed formally as the lowest-order approximation
in a systematic procedure sometimes referred to as adiabatic
elimination of fast variables \cite{CohenTannoudji} or Van Vleck
perturbation theory \cite{Hausinger}. The semiclassical
calculation treats the dynamical variables classically and is at
first sight not related to any specific approximation related to
the system parameters.

The weak-coupling approximation is suited for studying the
excitation-exchange dynamics between the qubit and oscillator, but
it does not give any hint of an instability in the system. The
main result of the adiabatically-adjusting-oscillator
approximation is the renormalized qubit gap. Apart from the
oscillatory behaviour in the gap, the Gaussian-function decrease
at large $\lambda/(\hbar\omega_0)$ values is a signature of the
strong entanglement between the qubit and the oscillator in the
energy eigenstates. Nevertheless, no `critical point', i.e.~a
point that is associated with a sudden change in any of the
effective qubit parameters (particularly the renormalized gap), is
obtained in that calculation. The adiabatically-adjusting-qubit
approximation predicts a reduced effective oscillator frequency
for weak coupling (and assuming that the qubit is in its ground
state), and a qualitative change in behaviour upon crossing the
critical point
\begin{equation}
\frac{4\lambda^2}{\hbar\omega_0 E_{\rm q}} = 1.
\end{equation}
Above the critical point, the energy eigenstates can be highly
entangled qubit-oscillator states. As in the
adiabatically-adjusting-oscillator approximation, the separation
between neighbouring energy levels is found to follow a
Gaussian-function dependence in the parameter
$\lambda/(\hbar\omega_0)$. The adiabatically-adjusting-oscillator
and adiabatically-adjusting-qubit approximations give different
predictions regarding the typical value of $\lambda$ at which the
Gaussian-function decrease in energy separation starts: the former
gives $\lambda\sim\hbar\omega_0$ and the latter gives
$\lambda\sim\sqrt{\hbar\omega_0\Delta}$. The semiclassical
calculation produces the same critical-point condition as the
adiabatically-adjusting-qubit approximation. Even though the
semiclassical calculation naturally cannot produce any
entangled-state solutions, its results can be used as a starting
point for studying quantum superpositions of the different
semiclassical solutions.

One could understand the reason for the absence of a critical
point in the case of a high-frequency oscillator as having to do
with the pairing of energy levels. In this case, the energy levels
form pairs all the way from $\lambda=0$ to
$\lambda\rightarrow\infty$. In contrast, in the case of a
high-frequency qubit the low-lying levels are equally spaced for
small values of $\lambda$. As $\lambda$ increases, the energy
levels start approaching each other while remaining equally
spaced, a situation that corresponds to a decreasing renormalized
oscillator frequency. At the point where the energy levels are
expected to collapse to a single, highly degenerate energy level,
they pair up and the different pairs start moving away from each
other. The energy levels now resemble those of an increasingly
deep double-well potential. Thus the energy levels and energy
eigenstates exhibit two qualitatively different structures below
and above the critical point.

It is worth mentioning that the adiabatically-adjusting-oscillator
and adiabatically-adjusting-qubit approximations start with
similar, or symmetric, reasoning. The asymmetry in the results is
mainly due to the different dependence in the energy levels and
energy eigenstates of the fast subsystem on the state of the slow
subsystem. In the case of a fast qubit, the qubit's energy
produces the largest effect on the slow oscillator. In the case of
a fast oscillator, the oscillator's energy does not depend on the
state of the qubit, and only the changes in the energy eigenstates
lead to effective changes to the behaviour of the slow qubit.

One might wonder why the results of the semiclassical calculation
agree with those of the high-frequency-qubit approximation. This
agreement can be understood by noting first that the oscillator
has continuous variables, such that it is conceivable that certain
states will be described to a good approximation using classical
variables (In this context one can think of coherent states, which
to a good approximation behave classically). When the qubit's
frequency is high, it is also conceivable that the qubit's state,
which for example follows the instantaneous ground state, can be
described by the classical variables that specify the
instantaneous ground state. In this case, one can expect the
semiclassical calculation to give good results. One can also use
this argument to conclude that a phase-transition-like singularity
will occur in the limit $\hbar\omega_0/E_{\rm q}\rightarrow 0$,
where the semiclassical calculation can be expected to give exact
results. In contrast, in the limit of a low-frequency qubit, the
dynamics will necessarily be described by the coupling of two
discrete quantum states, and this situation cannot be described
well using a semiclassical approximation.

In the context of discussing the phase-transition-like bifurcation
in this system, it is worth mentioning a related system with a
true phase transition: the Dicke model \cite{Dicke}. If one
replaces the single qubit by a large number of qubits with equal
values of $E_{\rm q}$, all coupled to the same cavity with the
same value of $\lambda$, then by taking the appropriate
thermodynamic limit ($N\rightarrow \infty$, $\lambda^2
N=\tilde{\lambda}^2$) one finds a phase transition between states
similar to those discussed above \cite{Hepp}. The critical point
is given by the condition $4\tilde{\lambda}^2=\hbar\omega_0 E_{\rm
q}$, in analogy to the critical-point condition discussed in
Secs.~\ref{Sec:AnalyticalMethodsAdiabaticQubit} and
\ref{Sec:AnalyticalMethodsSemiclassical}. In contrast to the
single-qubit case, however, the phase transition now occurs
regardless of the relation between the qubit and oscillator
frequencies. Note that the qubits behave collectively as a single
large spin in this case (and for low-lying states a large spin
behaves similarly to a harmonic oscillator), such that the entire
system can be approximated by two coupled oscillators. Note also
that the semiclassical calculation arises naturally in this case:
when the effective spin has an infinite number of allowed states,
it is natural to make a classical approximation where fluctuations
in the spin are small compared to the total size of the available
state space.

\section{Numerical calculations}
\label{Sec:NumericalCalculations}

In this section we present results of numerical calculations that
demonstrate the properties of the system in the different
parameter regimes. In particular, we perform calculations for the
resonant case, the high-frequency-oscillator case and the
high-frequency-qubit case. We also vary the qubit bias $\epsilon$,
or alternatively the angle $\theta$, in order to analyze its
effect on the properties of the system.

\subsection{Energy-level spectrum}
\label{Sec:NumericalCalculationsEnergyLevels}

\begin{figure}[h]
\includegraphics[width=7.0cm]{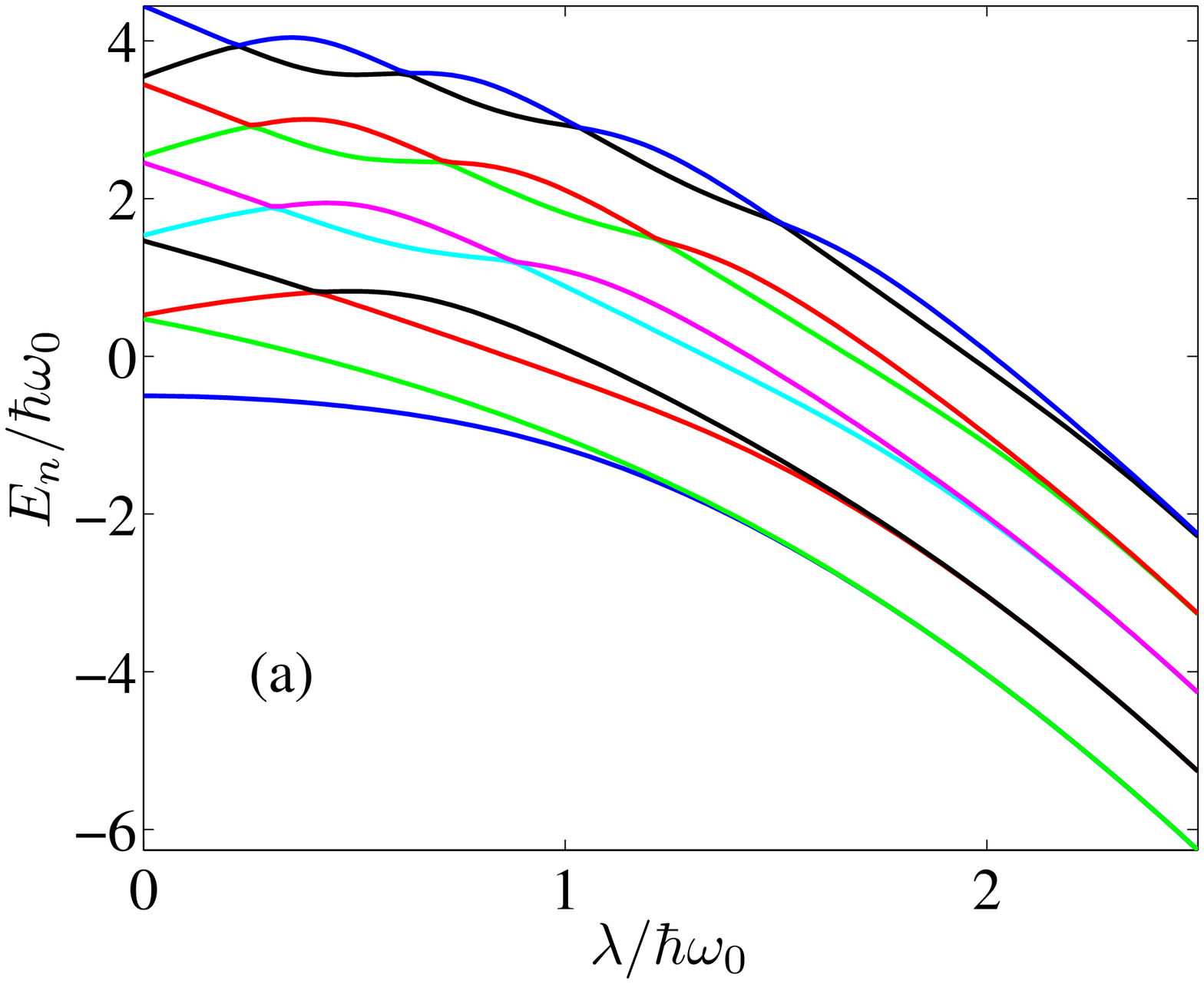}
\includegraphics[width=7.0cm]{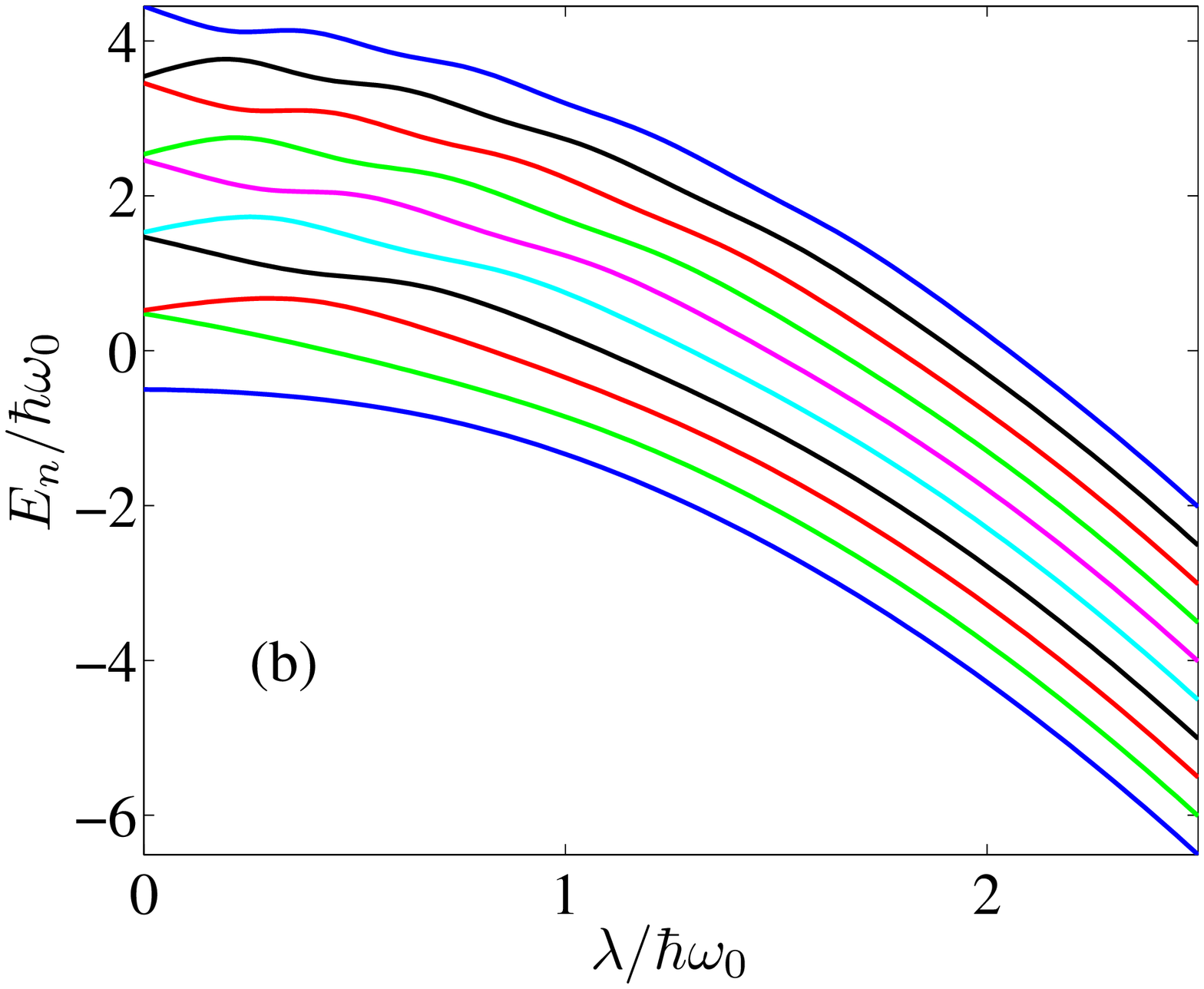}
\includegraphics[width=7.0cm]{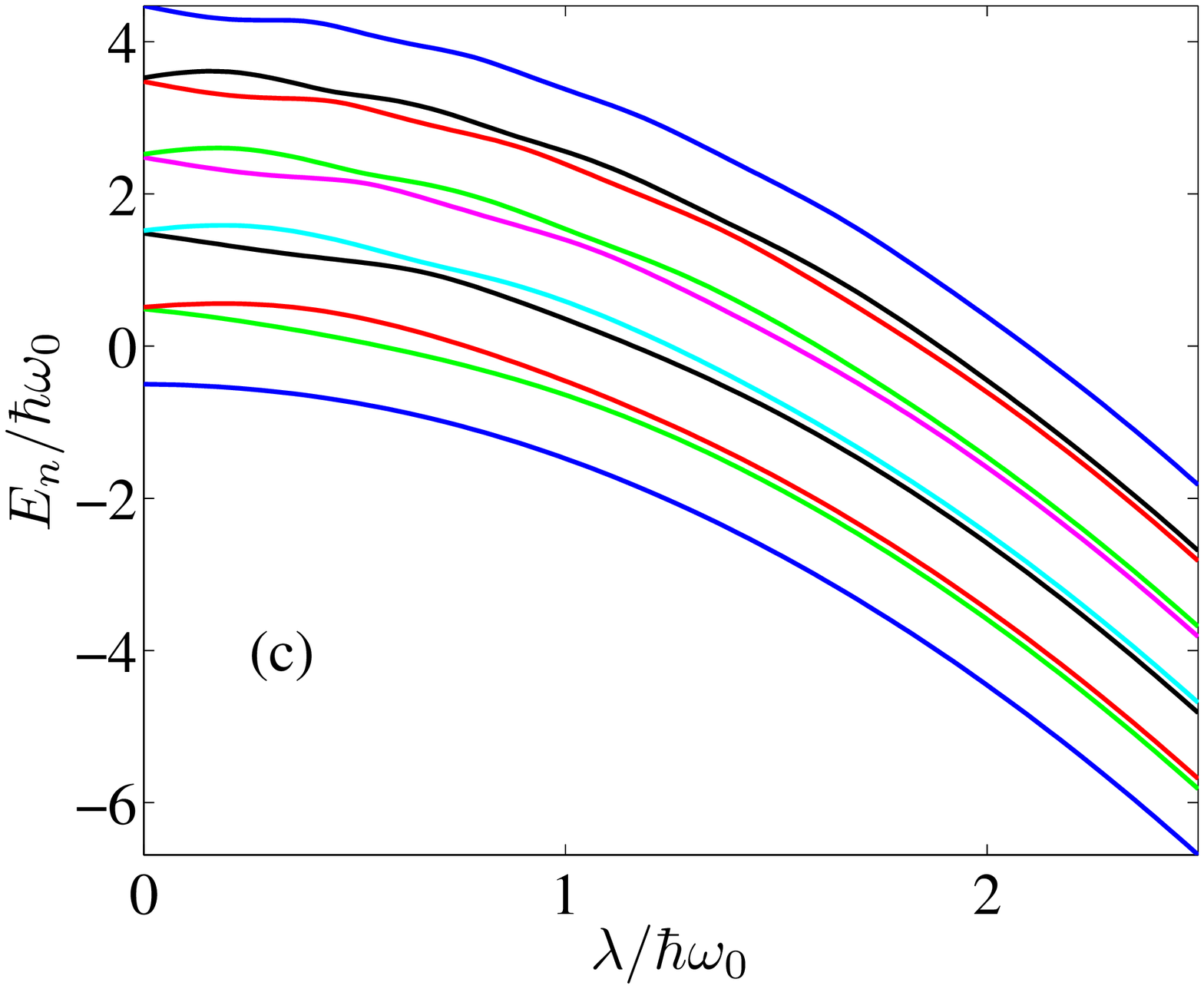}
\caption{(Color online) Lowest ten energy levels in the resonant
case, i.e.~when $\hbar\omega_0/E_{\rm q}=1$. The rescaled energy
$E_n/(\hbar\omega_0)$ with $n=1,2,...,10$ is plotted as a function
of the rescaled coupling strength $\lambda/(\hbar\omega_0)$.
Panels (a), (b) and (c) correspond to $\theta=0,\pi/6$ and
$\pi/3$, respectively [recall that
$\theta=\arctan(\epsilon/\Delta)$].}
\label{Fig:EnergyLevelsResonant}
\end{figure}

In Fig.~\ref{Fig:EnergyLevelsResonant} we plot the energies of the
lowest ten levels as a function of the coupling strength $\lambda$
in the resonant case $\hbar\omega_0= E_{\rm q}$. When
$\epsilon=\lambda=0$, the ground state is non-degenerate and each
higher energy level is doubly degenerate. The separation between
the levels is $\hbar\omega_0$, which is also equal to $E_{\rm q}$.
As $\lambda$ increases, the energy levels shift up or down, and
several avoided crossings are encountered. In the large $\lambda$
limit, all energy levels become doubly degenerate (i.e.~they form
pairs), including the ground state. The separation between the
different pairs of energy levels in this limit is again
$\hbar\omega_0$. These results agree with the picture of the
effective double-well potential of
Sec.~\ref{Sec:AnalyticalMethodsAdiabaticQubit}. For a small but
finite bias $\epsilon$ (i.e.~small but finite $\theta$) and small
coupling strength $\lambda$, the overall energy level structure is
similar to that in the $\epsilon=0$ case, except that the levels
do not approach each other as much at the avoided crossings. In
the large $\lambda$ limit, there are no degeneracies: the energy
levels are separated by the alternate distances $\epsilon$ and
$\hbar\omega_0-\epsilon$. This structure reflects the small
asymmetry in the double-well potential caused by a small tilt. For
large $\theta$ (i.e.~$\sin\theta\sim 1$), all features in the
spectrum are suppressed, except for the overall decrease in the
energy with increasing $\lambda$.

\begin{figure}[h]
\includegraphics[width=7.0cm]{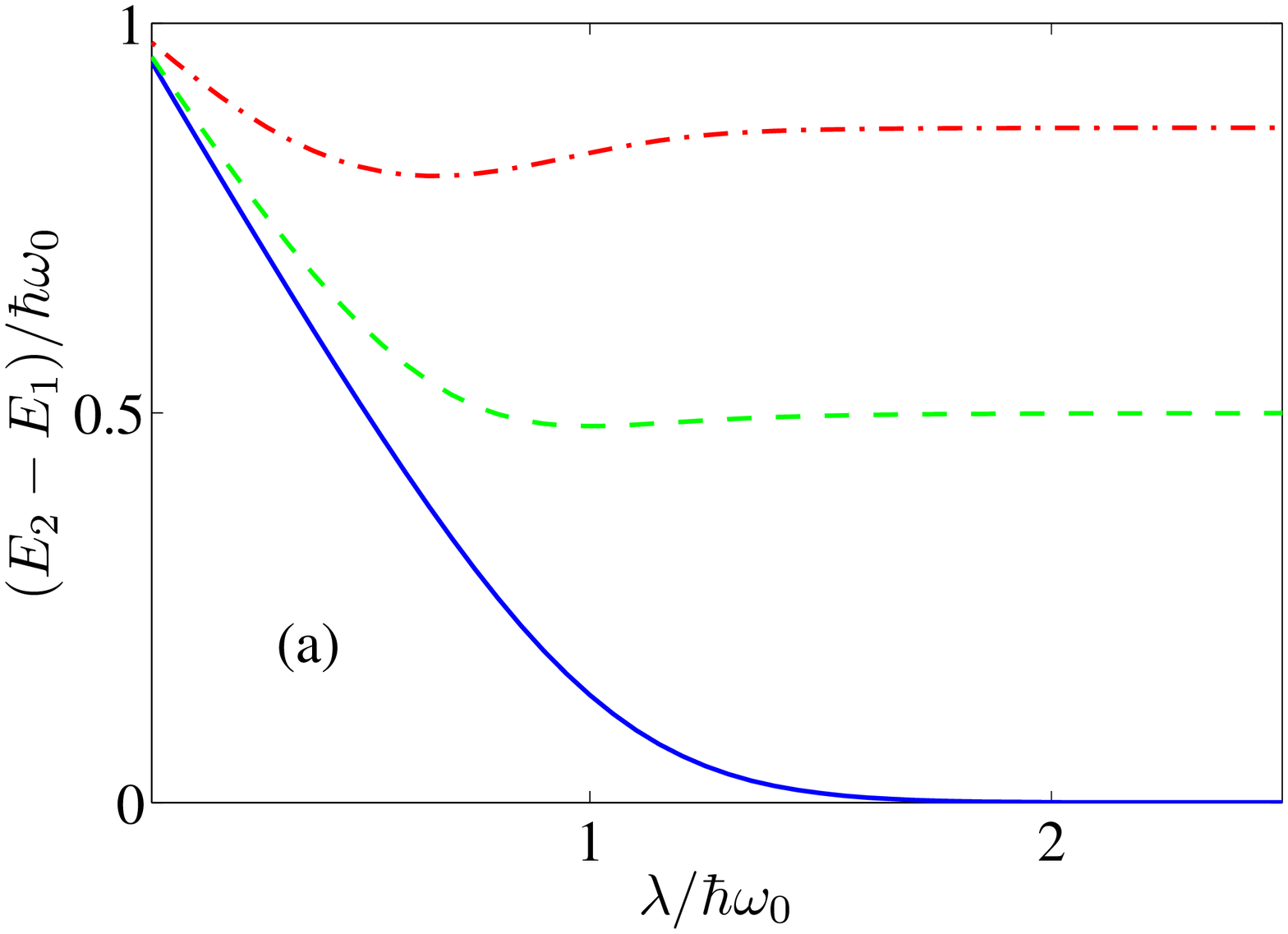}
\includegraphics[width=7.0cm]{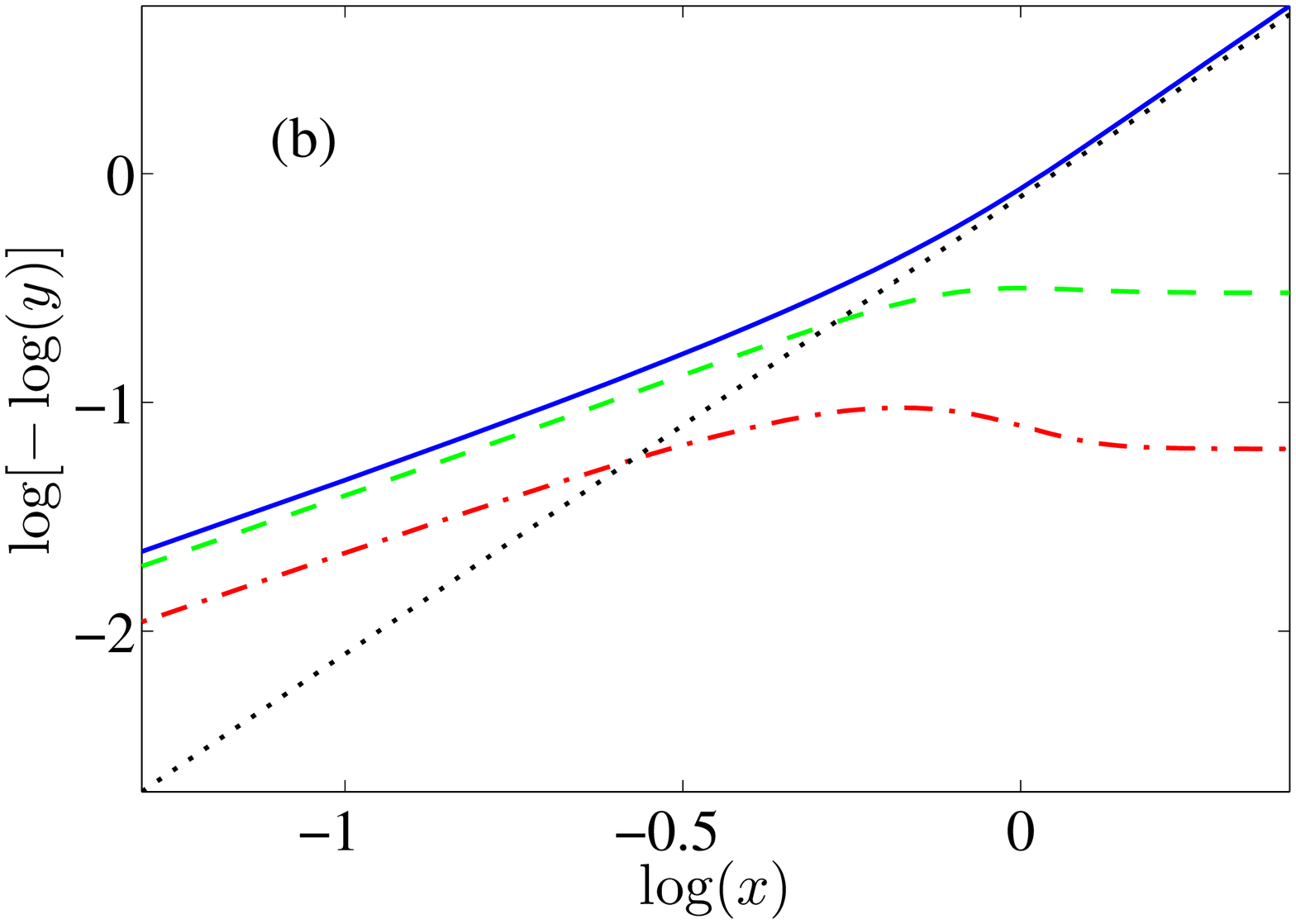}
\caption{(Color online) The separation between the lowest two
energy levels in the resonant case, i.e.~when
$\hbar\omega_0/E_{\rm q}=1$. In (a), the rescaled energy
separation $(E_2-E_1)/(\hbar\omega_0)$ is plotted as a function of
the rescaled coupling strength $\lambda/(\hbar\omega_0)$. The
blue, solid line corresponds to $\theta=0$; the green, dashed line
corresponds to $\theta=\pi/6$; and the red, dash-dotted line
corresponds to $\theta=\pi/3$. In (b), the same data is plotted on
a logarithmic scale in order to make a comparison with the formula
$E_2-E_1 = \Delta \exp \{-2(\lambda/\hbar\omega_0)^2\}$ from
Eq.~(\ref{Eq:Renormalized_gap}): $x$ and $y$ in the axis labels
refer to the axis labels in (a). The black, dotted line shows the
asymptotic behavior of the above formula. The good fit between the
blue and black lines means that the numerical results agree with
the results of Sec.~\ref{Sec:AnalyticalMethods} (We could extend
the range of agreement by plotting
$\log[\log(\Delta/\hbar\omega_0)-\log(y)]$; however, we are mostly
interested in demonstrating the agreement for large values of $x$,
where this modification would have little effect on the shape of
the blue curve). Similar figures can be generated for the other
values of $\hbar\omega_0/E_{\rm q}$. However, we do not show such
figures here.} \label{Fig:EnergyLevelsResonantPairSeparation}
\end{figure}

In order to examine the strong-coupling limit more closely, in
Fig.~\ref{Fig:EnergyLevelsResonantPairSeparation} we plot the
energy-level separation between the lowest two energy levels. The
results agree with the predictions of
Eq.~(\ref{Eq:Renormalized_gap}): deep in the strong-coupling
regime, the separation within the pairs of energy levels is given
by $E_{2n+2}-E_{2n+1} \sim (\lambda/\hbar\omega_0)^n \exp
\{-2(\lambda/\hbar\omega_0)^2\}$ for $\epsilon=0$ and by
$\epsilon$ for $\epsilon\neq 0$.

\begin{figure}[h]
\includegraphics[width=7.0cm]{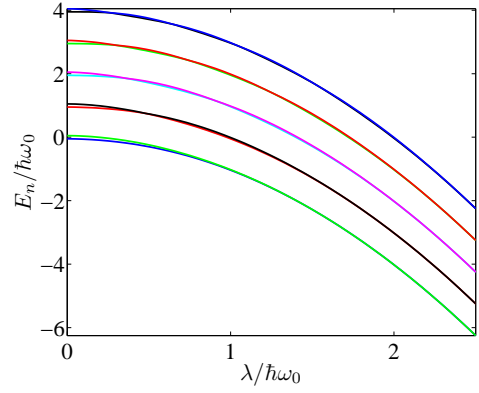}
\caption{(Color online) Lowest ten energy levels in the case of a
high-frequency oscillator; $\hbar\omega_0/E_{\rm q}=10$. Here we
only show the results for $\theta=0$, because the overall
appearance of the plots is independent of $\theta$. More details
can be seen in
Fig.~\ref{Fig:EnergyLevelsLargeOmegaPairSeparation}.}
\label{Fig:EnergyLevelsLargeOmega}
\end{figure}

\begin{figure}[h]
\includegraphics[width=7.0cm]{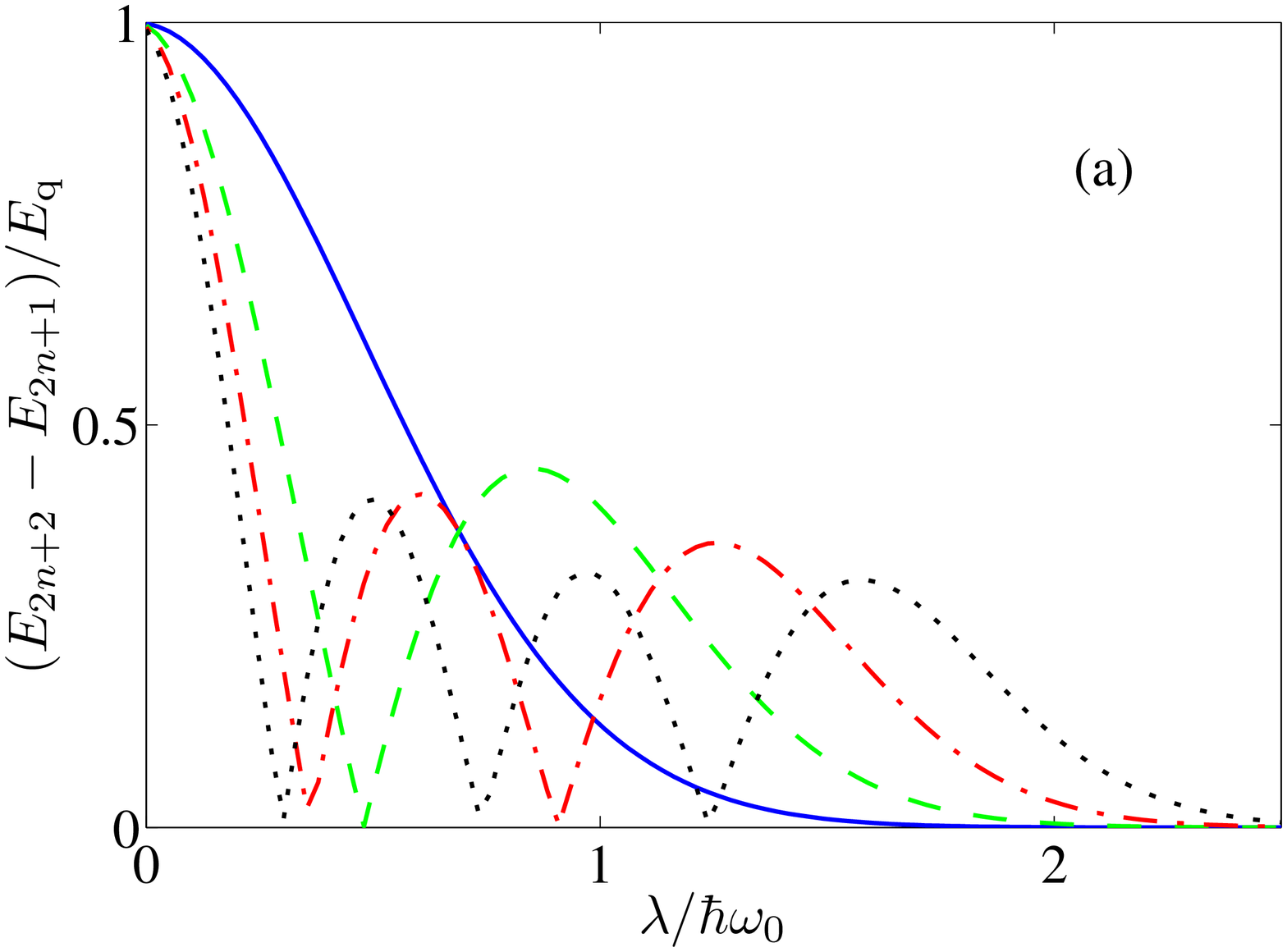}
\includegraphics[width=7.0cm]{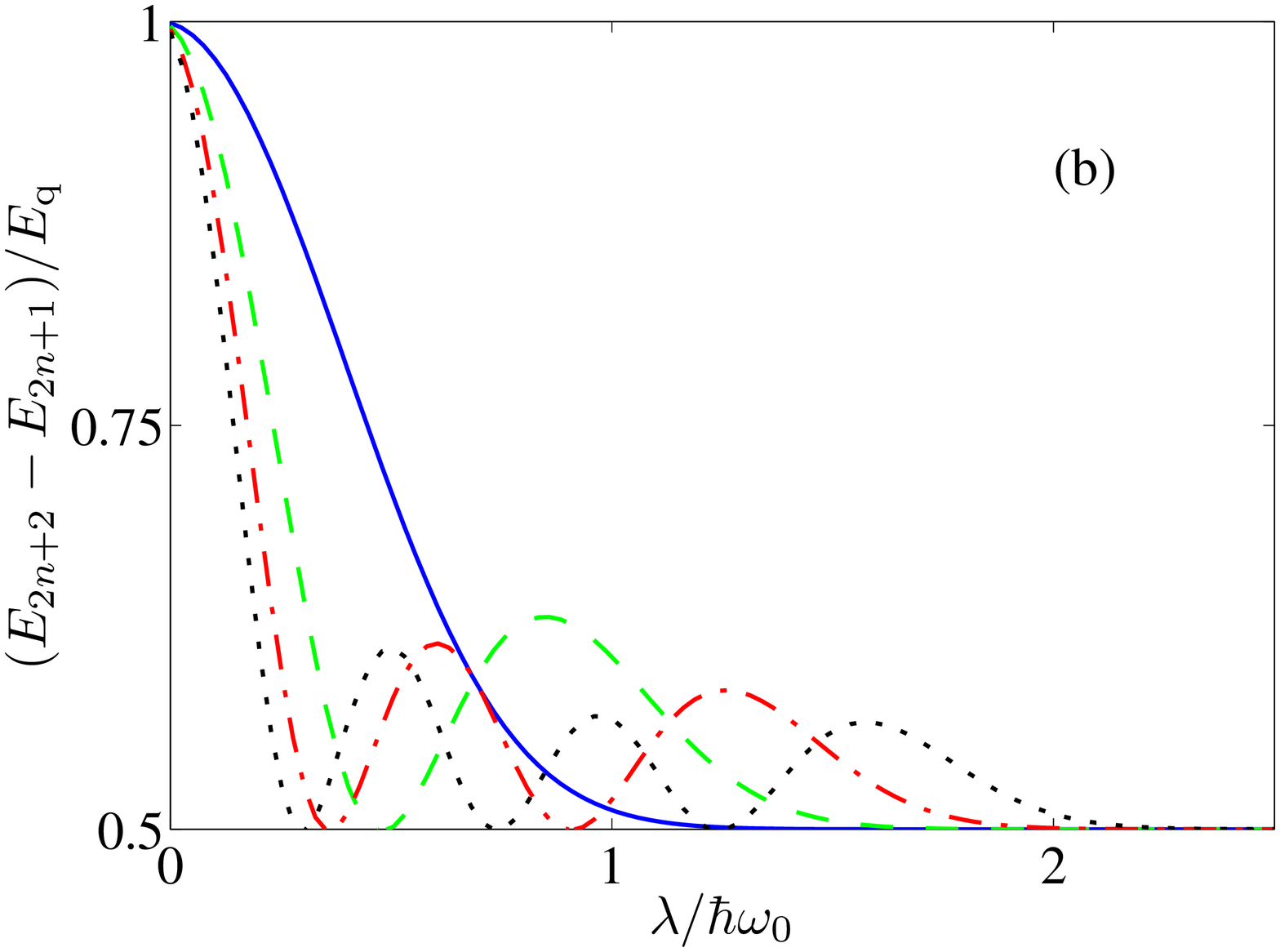}
\includegraphics[width=7.0cm]{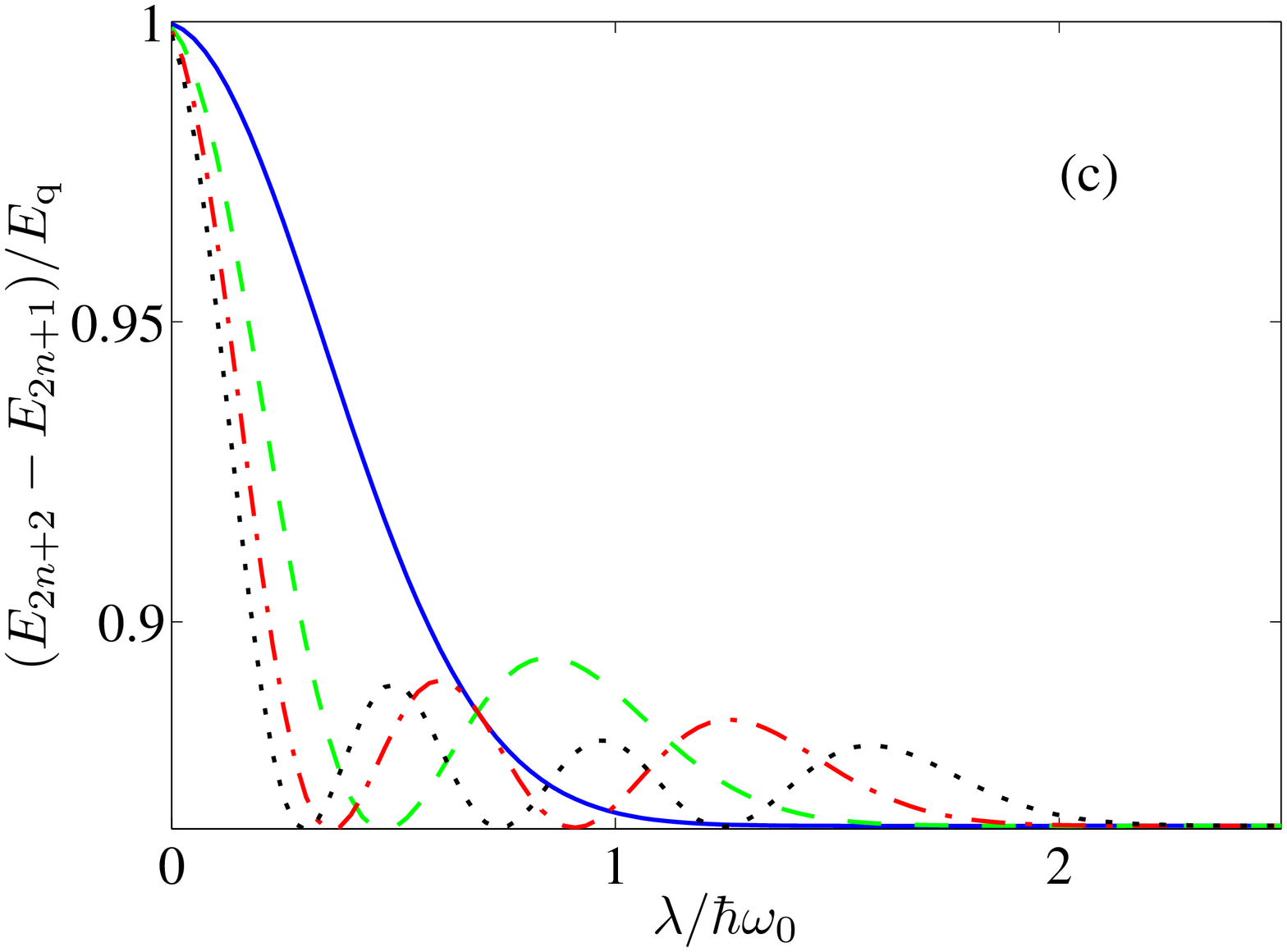}
\caption{(Color online) The rescaled energy separation
$(E_{2n+2}-E_{2n+1})/E_{\rm q}$ within the lowest four pairs of
energy levels [i.e.~for $n=0$ (blue, solid line), 1 (green, dashed
line), 2 (red, dash-dotted line) and 3 (black, dotted line)], as a
function of $\lambda/(\hbar\omega_0)$. As in
Fig.~\ref{Fig:EnergyLevelsLargeOmega}, we take
$\hbar\omega_0/E_{\rm q} = 10$. In panels (a), (b) and (c),
$\theta=0,\pi/6$ and $\pi/3$, respectively. Note that the minimum
value on the $y$-axis is given by $\sin\theta$ and is different in
the three panels.}
\label{Fig:EnergyLevelsLargeOmegaPairSeparation}
\end{figure}

In Fig.~\ref{Fig:EnergyLevelsLargeOmega} we plot the energies of
the lowest ten levels as a function of $\lambda$ in the case of a
high-frequency oscillator, i.e.~when $E_{\rm q} \ll
\hbar\omega_0$. As explained in
Sec.~\ref{Sec:AnalyticalMethodsAdiabaticOscillator}, if one
considers a pair of energy levels, e.g.~the lowest two energy
levels, one has a modified effective qubit Hamiltonian. When
$\lambda=0$, one recovers the bare qubit Hamiltonian. As $\lambda$
increases, the effective qubit gap $\Delta$ decreases and
approaches zero in the limit $\lambda/(\hbar\omega_0) \rightarrow
\infty$. In Fig.~\ref{Fig:EnergyLevelsLargeOmegaPairSeparation} we
plot the separations within the four lowest pairs of energy
levels. The effective gap follows the shape of a Gaussian function
times a Laguerre polynomial, vanishing at the zeros of the
Laguerre polynomial. As $\theta$ is increased from zero, i.e. as
the ratio $\Delta/\epsilon$ decreases, the dependence of the
energy-level separation on the coupling strength becomes weaker
(This phenomenon can be seen by comparing the different panels in
Fig.~\ref{Fig:EnergyLevelsLargeOmegaPairSeparation}). Note that
the location of the peaks does not change, but the effect of the
gap on the energy levels becomes smaller with increasing $\theta$.

\begin{figure}[h]
\includegraphics[width=7.0cm]{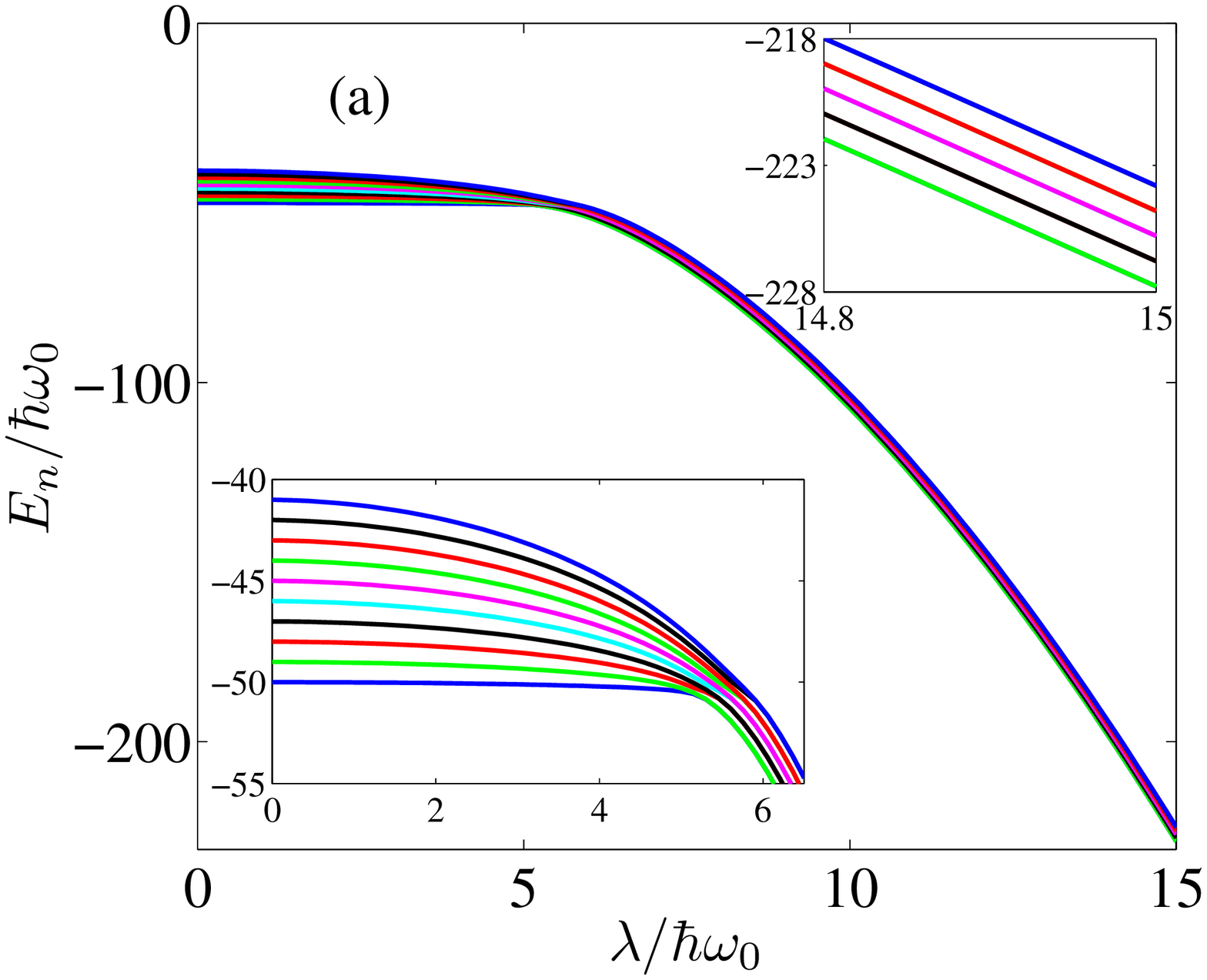}
\includegraphics[width=7.0cm]{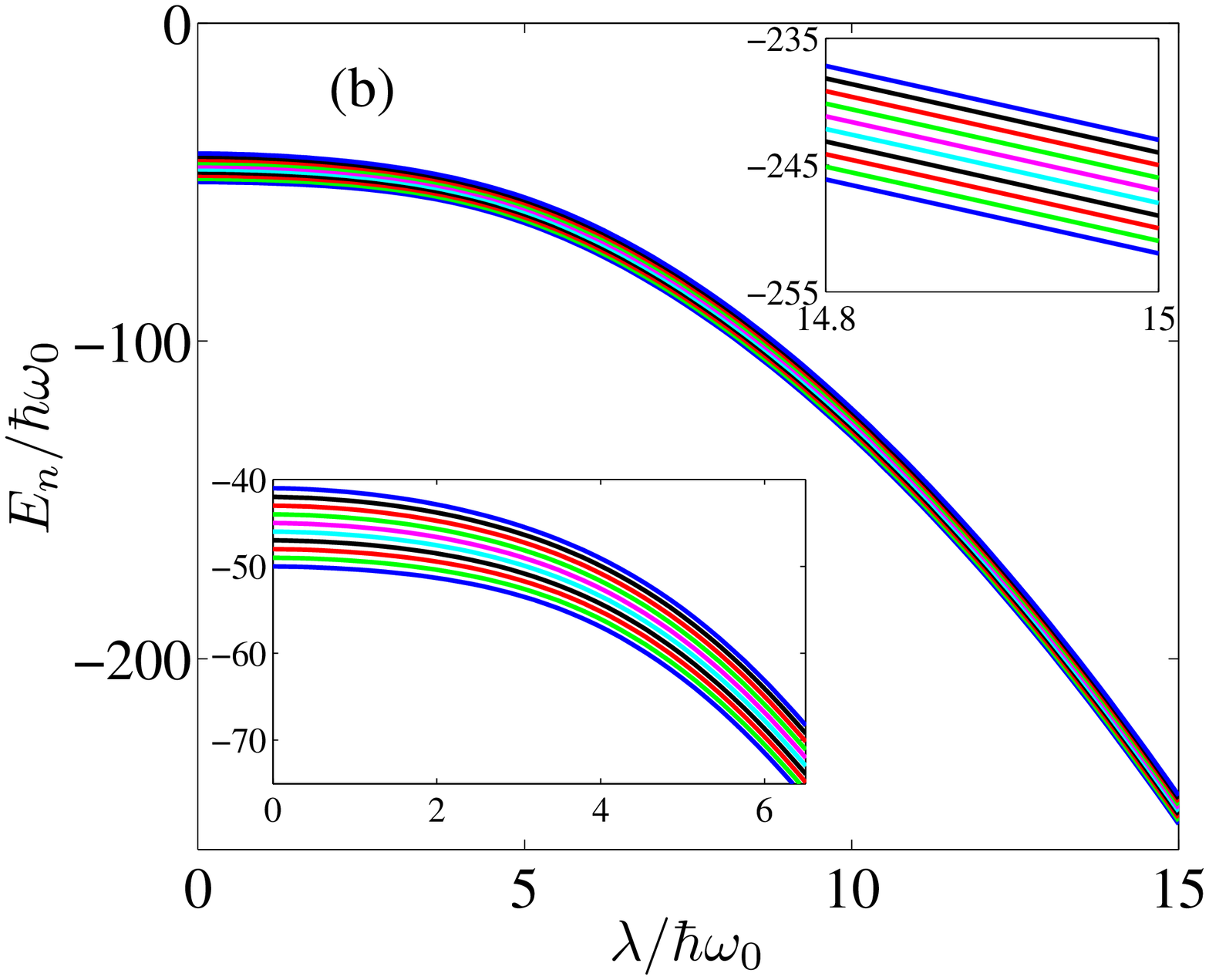}
\includegraphics[width=7.0cm]{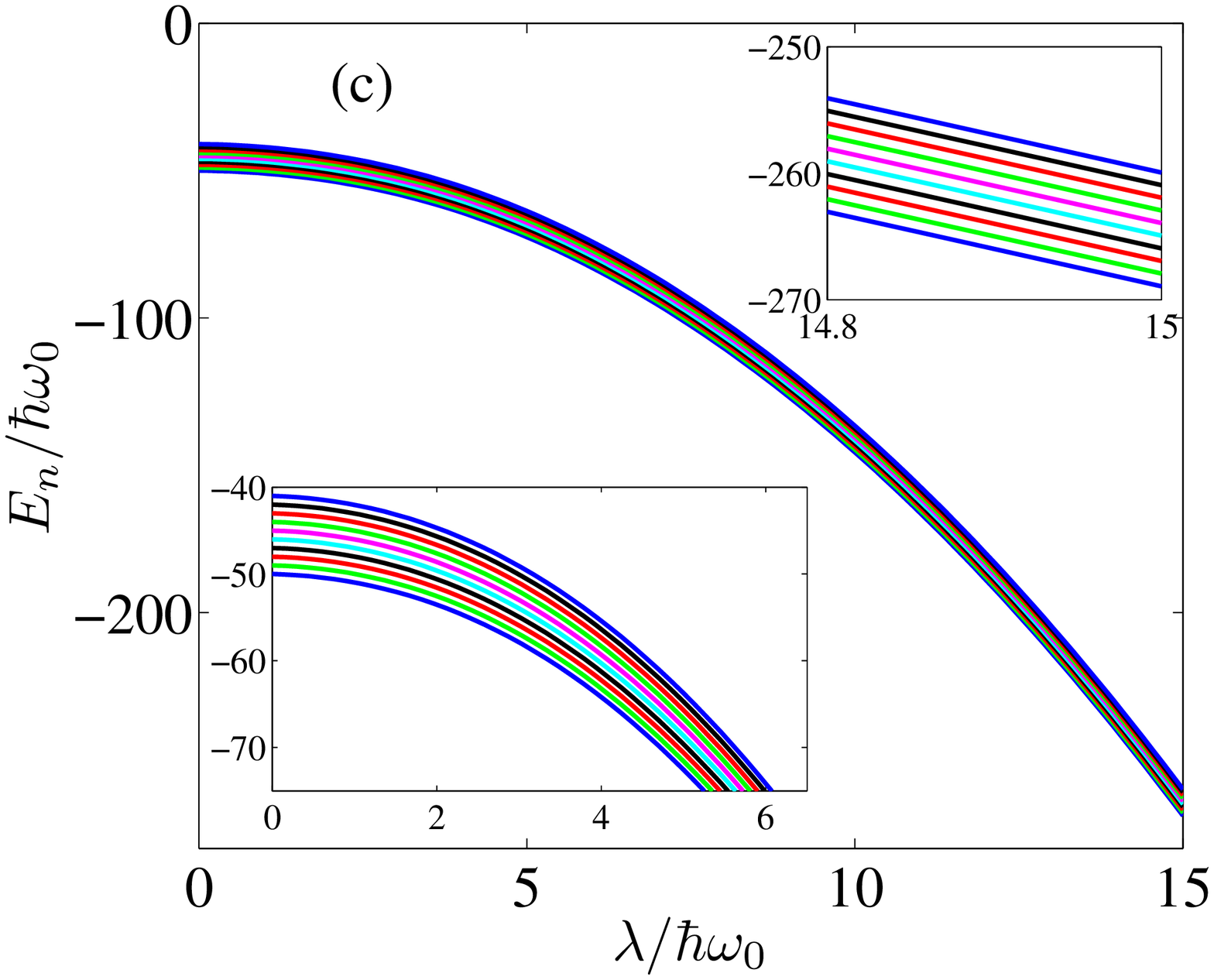}
\caption{(Color online) Lowest ten energy levels in the case of a
high-frequency qubit; $\hbar\omega_0/E_{\rm q}=0.01$. In panels
(a), (b) and (c), $\theta=0,\pi/6$ and $\pi/3$, respectively. The
insets show enlarged views of the weak-coupling and
strong-coupling regions.} \label{Fig:EnergyLevelsLargeDelta}
\end{figure}

In Fig.~\ref{Fig:EnergyLevelsLargeDelta} we plot the energies of
the lowest ten levels as a function of $\lambda$ in the case of a
high-frequency qubit, i.e.~when $E_{\rm q} \gg \hbar\omega_0$. The
most dramatic effects occur for $\theta=0$. The ground-state
energy remains essentially constant between $\lambda=0$ and
$\lambda=\sqrt{\hbar\omega_0\Delta}/2$. Beyond this point the
ground-state energy decreases indefinitely with increasing
$\lambda$. Furthermore, below the critical point, the low-lying
energy levels approach each other with increasing $\lambda$ as if
they were going to collapse to one point, as would be expected for
a vanishing $\tilde{\omega}_0$. Above the critical point, the
energy levels form pairs whose intra-pair separation decreases
with increasing $\lambda$. The above scenario is suppressed as
$\theta$ is increased. There is no longer any sign of a critical
point, and the energy-level separations are independent of
$\lambda$.

\subsection{Squeezing, entanglement, and `cat-ness' in the ground state}
\label{Sec:NumericalCalculationsEntanglementAndSqueezing}

One obvious possibility for the preparation of squeezed, entangled
or Schr\"odinger-cat states in the case of ultrastrong coupling is
to have a ground state that exhibits one of these unusual
properties. With this point in mind, in this section we analyze
the oscillator's squeezing and cat-ness as well as the
qubit-oscillator entanglement in the ground state for different
choices of system parameters.

\begin{figure}[h!]
\includegraphics[width=7.5cm]{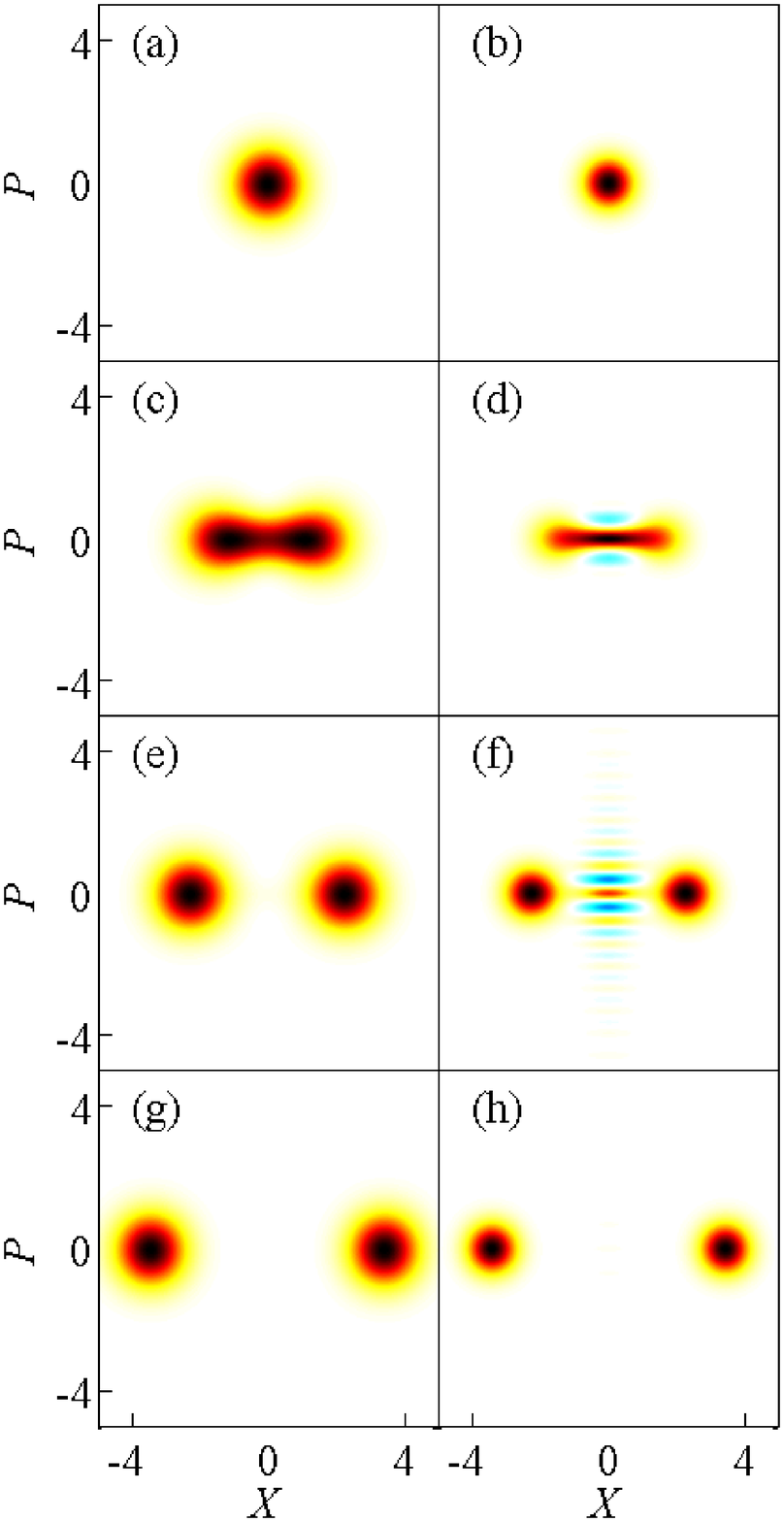}
\caption{(Color online) The $Q$ function (left) and the Wigner
function (right) of the oscillator's state in the ground state of
the combined system. Here we take $\hbar\omega_0/\Delta=0.1$ and
$\epsilon=0$. The different panels correspond to
$\lambda/(\hbar\omega_0)=0.5$ (a,b), 2 (c,d), 2.5 (e,f) and 3.5
(g,h). For clarity, we adjust the color scheme in the different
panels such that the highest point is always black. The red and
yellow colors also correspond to positive values. The white color
corresponds to zero value. The blue color represents negative
values of the Wigner function. The oscillator's state goes from a
coherent state with no photons (i.e.~the vacuum state) in the
absence of coupling, to a squeezed state for low to moderate
coupling strengths and then to a qubit-oscillator entangled state
for very strong coupling. Note that the state in panels g and h is
highly nonclassical, in particular highly entangled, even though
this fact cannot be seen in the $Q$ and Wigner functions.}
\label{Fig:Qfunctions}
\end{figure}

As a first step, we plot the $Q$ function and the Wigner function
of the oscillator's state in the ground state of the coupled
system. The $Q$ function is given by
\begin{equation}
Q(X,P) = \frac{1}{\pi} \bra{X+iP} \rho_{\rm osc} \ket{X+iP},
\end{equation}
where $\rho_{\rm osc}$ is the oscillator's reduced density matrix
after tracing out the qubit from the ground state, $\rho_{\rm
osc}={\rm Tr}_{\rm q}\{ \ket{\Psi_{\rm GS}} \bra{\Psi_{\rm GS}}
\}$ with $\ket{\Psi_{\rm GS}}$ being the ground state of the
combined system, and the bra and ket in the above formula are
coherent states:
\begin{equation}
\ket{\alpha} = \exp\{ \alpha \hat{a}^{\dagger} -\alpha^* \hat{a}
\} \ket{0}.
\end{equation}
The state $\ket{0}$ represents the vacuum state with the
oscillator in its ground state. The Wigner function is given by
\begin{equation}
W(X,P) = \frac{1}{2\pi\hbar} \int_{-\infty}^{\infty}
\bra{X+\frac{1}{2}X'} \rho_{\rm osc} \ket{X-\frac{1}{2}X'}
e^{iPX'} dX',
\end{equation}
where the bra and ket are now eigenstates of the position operator
$\hat{x}$, i.e.~they are highly localized in configuration space.
The $Q$ and Wigner functions for a sequence of $\lambda$ values
are shown in Fig.~\ref{Fig:Qfunctions}.

\begin{figure}[h!]
\includegraphics[width=7.0cm]{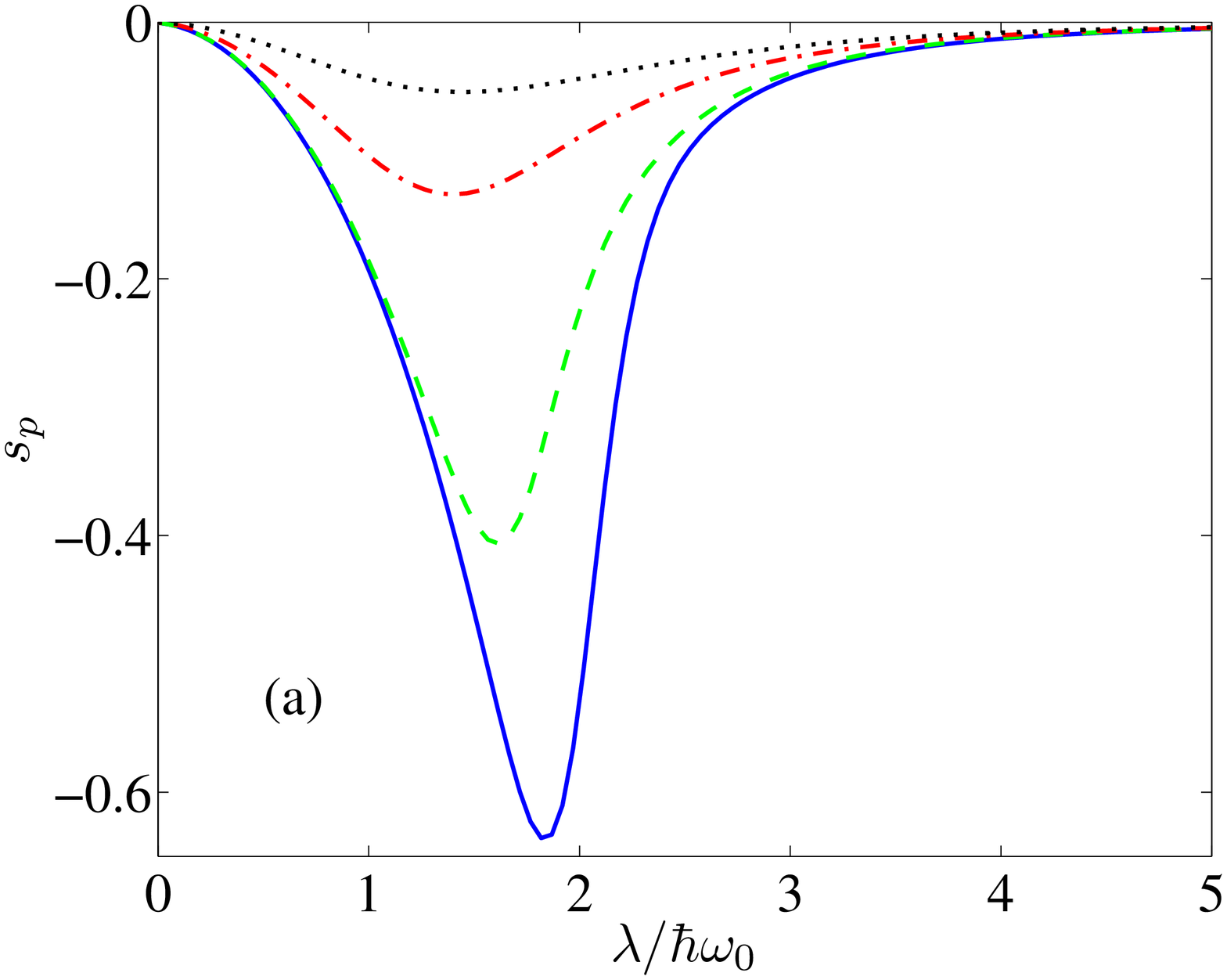}
\includegraphics[width=7.0cm]{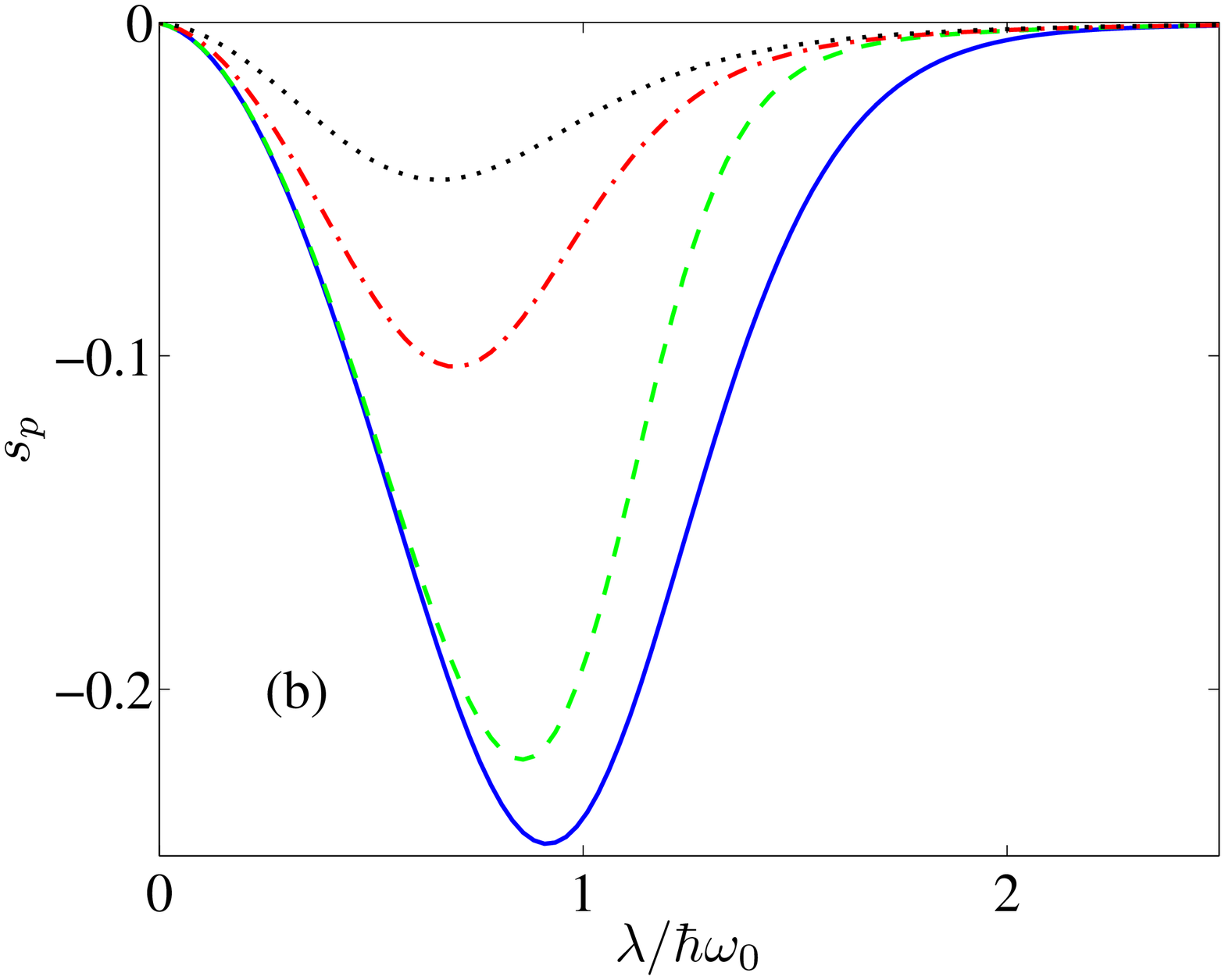}
\includegraphics[width=7.0cm]{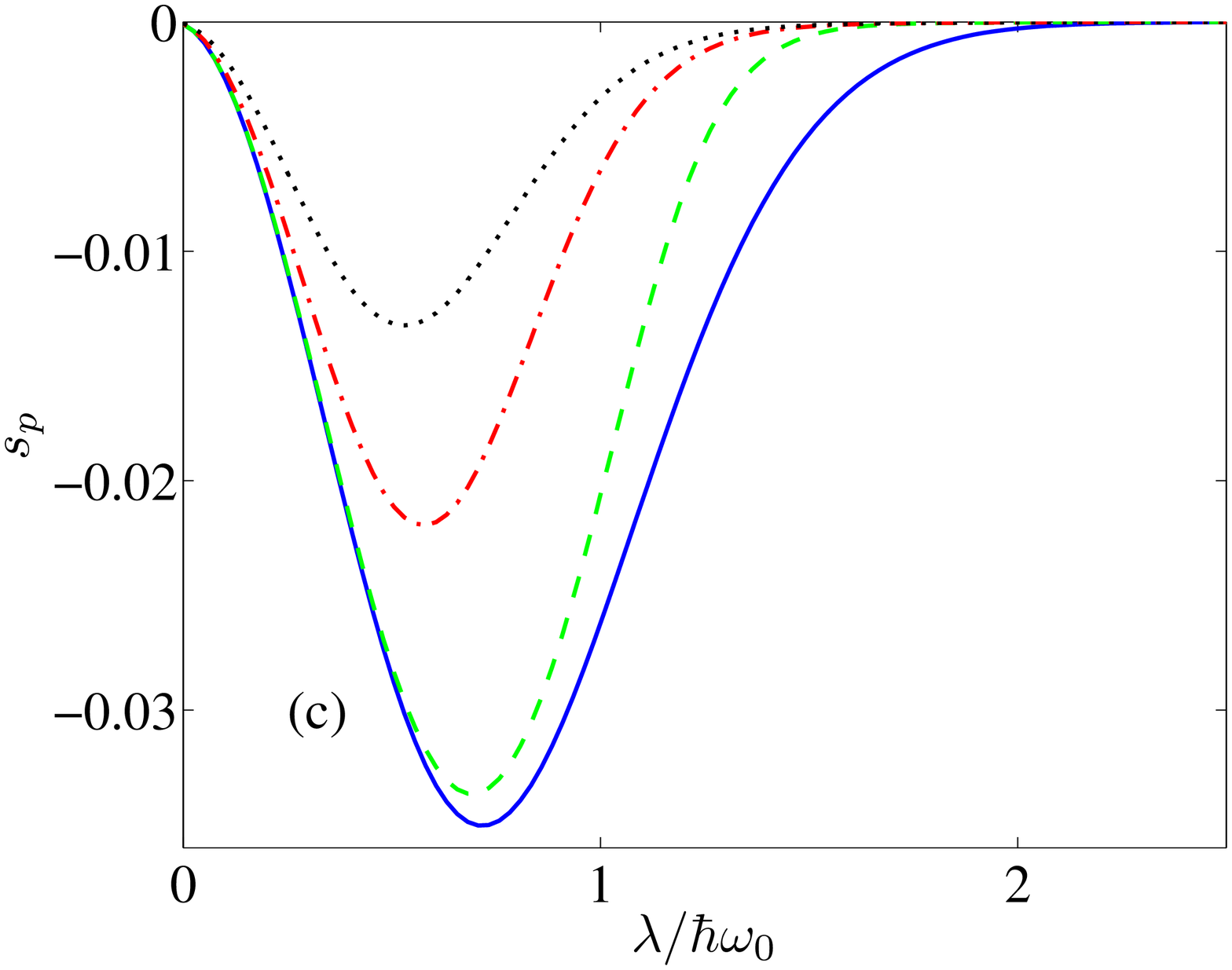}
\caption{(Color online) The momentum squeezing parameter $s_p$ as
a function of $\lambda/(\hbar\omega_0)$ for
$\hbar\omega_0/\Delta=0.1$ (a), 1 (b) and 10 (c). The different
curves correspond to $\epsilon/\Delta=0$ (blue, solid line), 0.1
(green, dashed line), 0.5 (red, dash-dotted line) and 1 (black,
dotted line). The oscillator's state becomes squeezed as the
coupling strength $\lambda$ increases, but then it reaches a
maximum and goes back to zero as the qubit and oscillator get
entangled in the strong-coupling regime. Note that the maximum
achievable squeezing decreases with increasing
$\hbar\omega_0/\Delta$.} \label{Fig:Squeezing}
\end{figure}

Going beyond the pictorial description shown above, a quantifier
for both the squeezing and cat-ness of the oscillator's state is
the set of two squeezing parameters in the $x$ and $p$ quadratures
as well as the product of the quadrature variances (Note here that
the oscillator's state is always mirror-symmetric with respect to
the $x$ axis in the setup under consideration, giving
$\langle\hat{p}\rangle=0$). After the appropriate conversion into
dimensionless variables, these quantifiers are given by
\begin{eqnarray}
s_x & = & 4 \left\langle \left( \hat{X} - \langle \hat{X} \rangle
\right)^2 \right\rangle - 1 \nonumber
\\
s_p & = & 4 \left\langle \left( \hat{P} - \langle \hat{P} \rangle
\right)^2 \right\rangle - 1 \nonumber
\\
K & = & \left\langle \left( \hat{x} - \langle \hat{x} \rangle
\right)^2 \right\rangle \left\langle \left( \hat{p} - \langle
\hat{p} \rangle \right)^2 \right\rangle \nonumber
\\
& = & \frac{\hbar^2}{4} (1+ s_x) (1+s_p).
\end{eqnarray}
The parameter $K$ is equal to $\hbar^2/4$ for a
minimum-uncertainty state (including both coherent and
quadrature-squeezed states) and is larger than that lower bound
for any other state (including Schr\"odinger-cat and
qubit-oscillator entangled states). In Fig.~\ref{Fig:Squeezing} we
plot the momentum squeezing parameter as a function of the
coupling strength $\lambda$. For small values of $\lambda$, the
squeezing increases with increasing $\lambda$. However, as
$\lambda$ increases further and the ground state becomes more and
more entangled, the squeezing is lost. The maximum achievable
squeezing is largest for the case of a high-frequency qubit,
$\hbar\omega_0\ll\Delta$. Indeed, as explained in
Sec.~\ref{Sec:AnalyticalMethodsAdiabaticQubit}, the oscillator's
effective potential becomes flatter and flatter as one approaches
the critical point, leading to a momentum squeezing parameter
close to $-1$.

As $|s_p|$ increases, one can ask whether the oscillator's state
is a quadrature-squeezed, minimum-uncertainty state or it deviates
from this ideal squeezed state. The answer to this question can be
obtained by analyzing the parameter $K$. We do not show any plots
of this parameter here. The main results are as follows: For the
case $\epsilon=0$, $K$ increases slowly and remains close to
$\hbar^2/4$ as $s_p$ increases, but near the maximum squeezing
point, $K$ starts increasing rapidly and diverges for
$\lambda/(\hbar\omega_0)\rightarrow\infty$. For finite values of
$\epsilon$, $K$ increases slightly above $\hbar^2/4$, but then
turns and goes back to $\hbar^2/4$ as $s_p$ goes back to zero in
the strong-coupling limit.

We have seen that squeezed states are obtained for weak to
moderate coupling. The question now is what states we have for
strong coupling. The $Q$ functions and the $s_p$ and $K$ results
discussed above do not distinguish between a Schr\"odinger-cat
state in the oscillator and a qubit-oscillator entangled state.
The Wigner function has negative values for moderately strong
coupling (Figs.~\ref{Fig:Qfunctions}d and \ref{Fig:Qfunctions}f),
indicating nonclassical states of the Schr\"odinger-cat type (Note
that quadrature-squeezed states have nonnegative Wigner
functions). In order to distinguish more clearly between
Schr\"odinger-cat states in the oscillator and qubit-oscillator
entangled states, we now analyze the entanglement properties in
the ground state.

The entanglement is quantified by the entropy $S$ of the qubit's
state. This quantity is obtained by calculating the ground state
of the combined system $\ket{\Psi_{\rm GS}}$, using it to obtain
the qubit's reduced density matrix in the ground state $\rho_{\rm
q}={\rm Tr}_{\rm osc}\{ \ket{\Psi_{\rm GS}} \bra{\Psi_{\rm GS}}
\}$, and then evaluating the entropy of that state $S=-{\rm Tr}\{
\rho_{\rm q} \log_2 \rho_{\rm q} \}$.

\begin{figure}[h!]
\includegraphics[width=7.0cm]{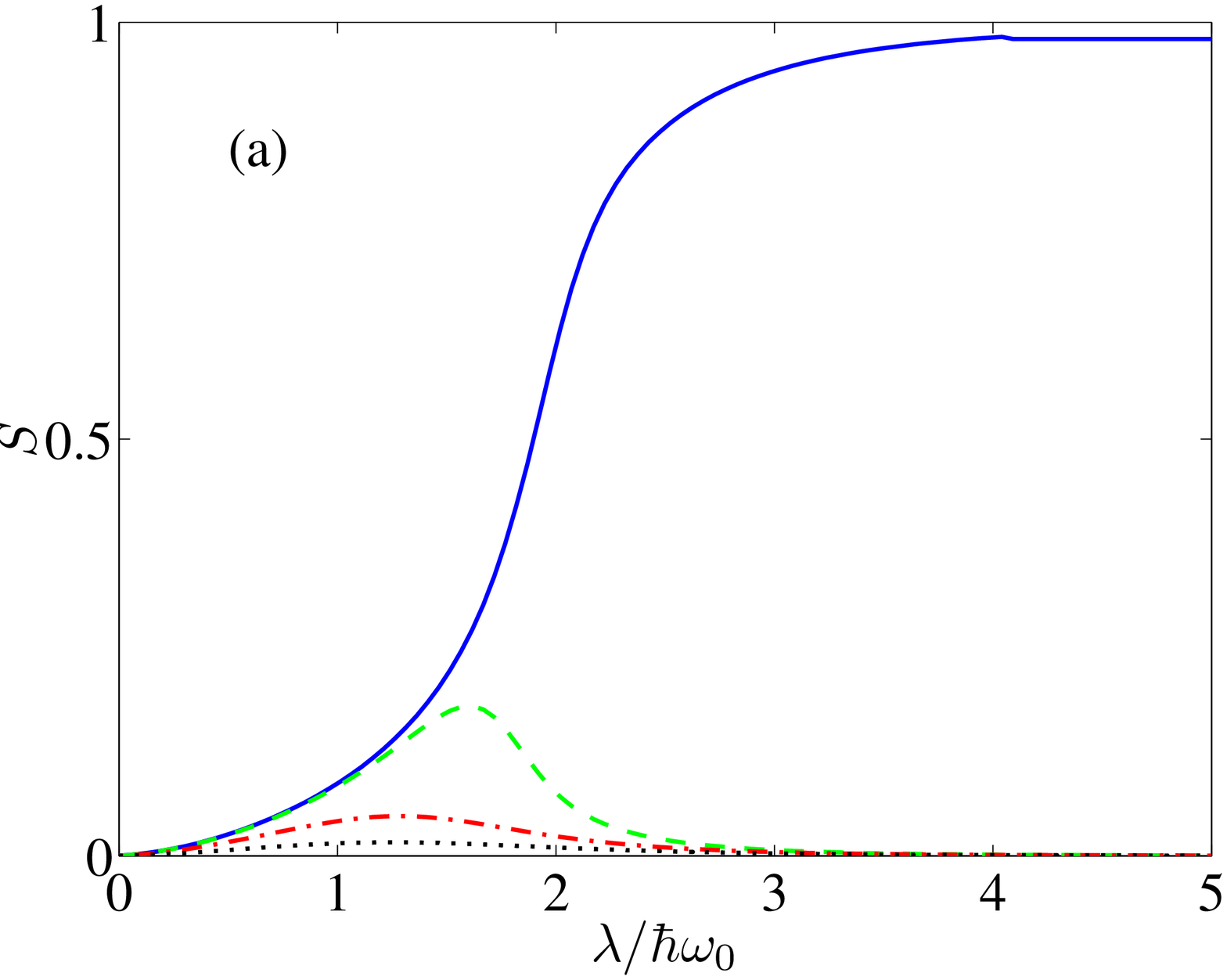}
\includegraphics[width=7.0cm]{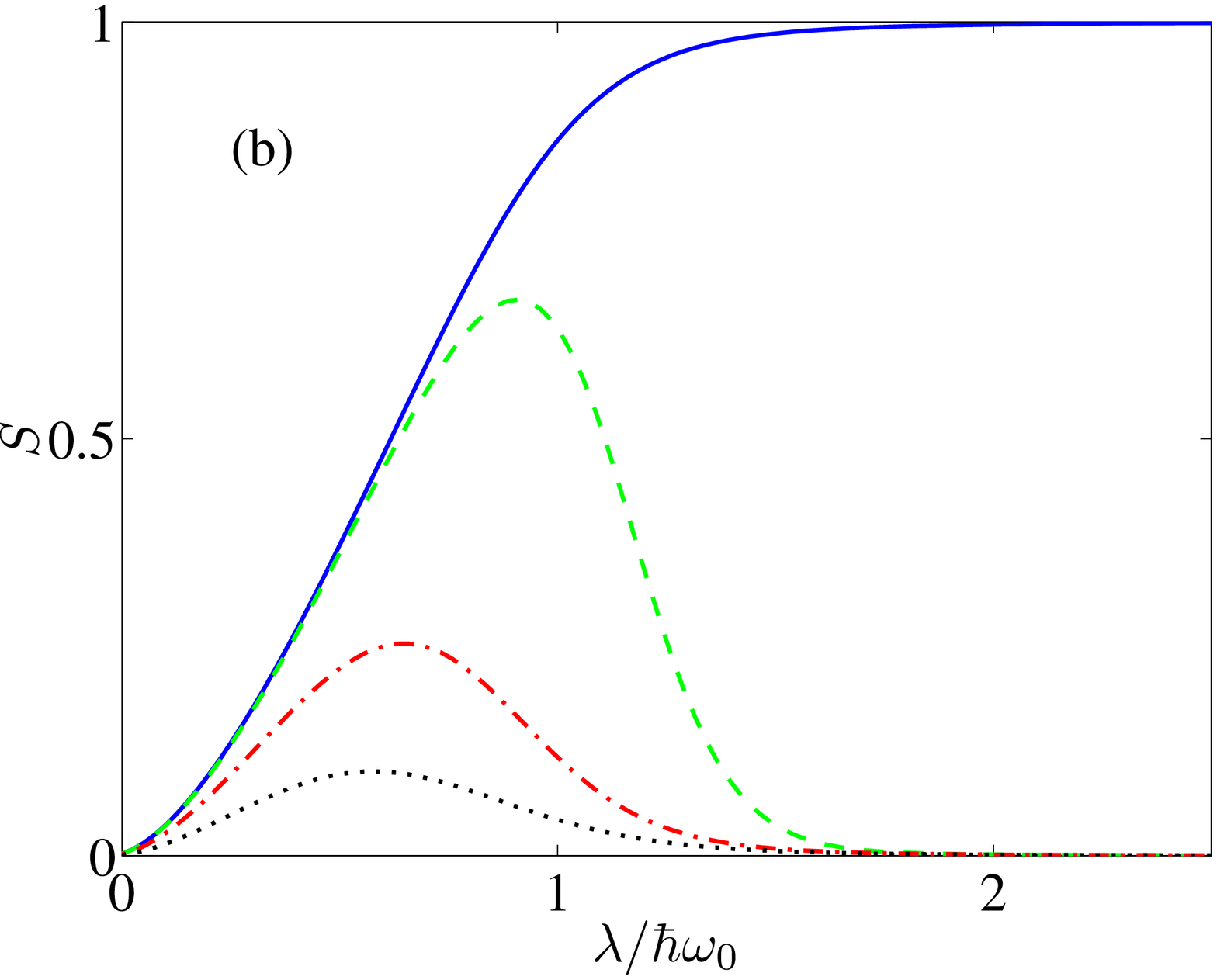}
\includegraphics[width=7.0cm]{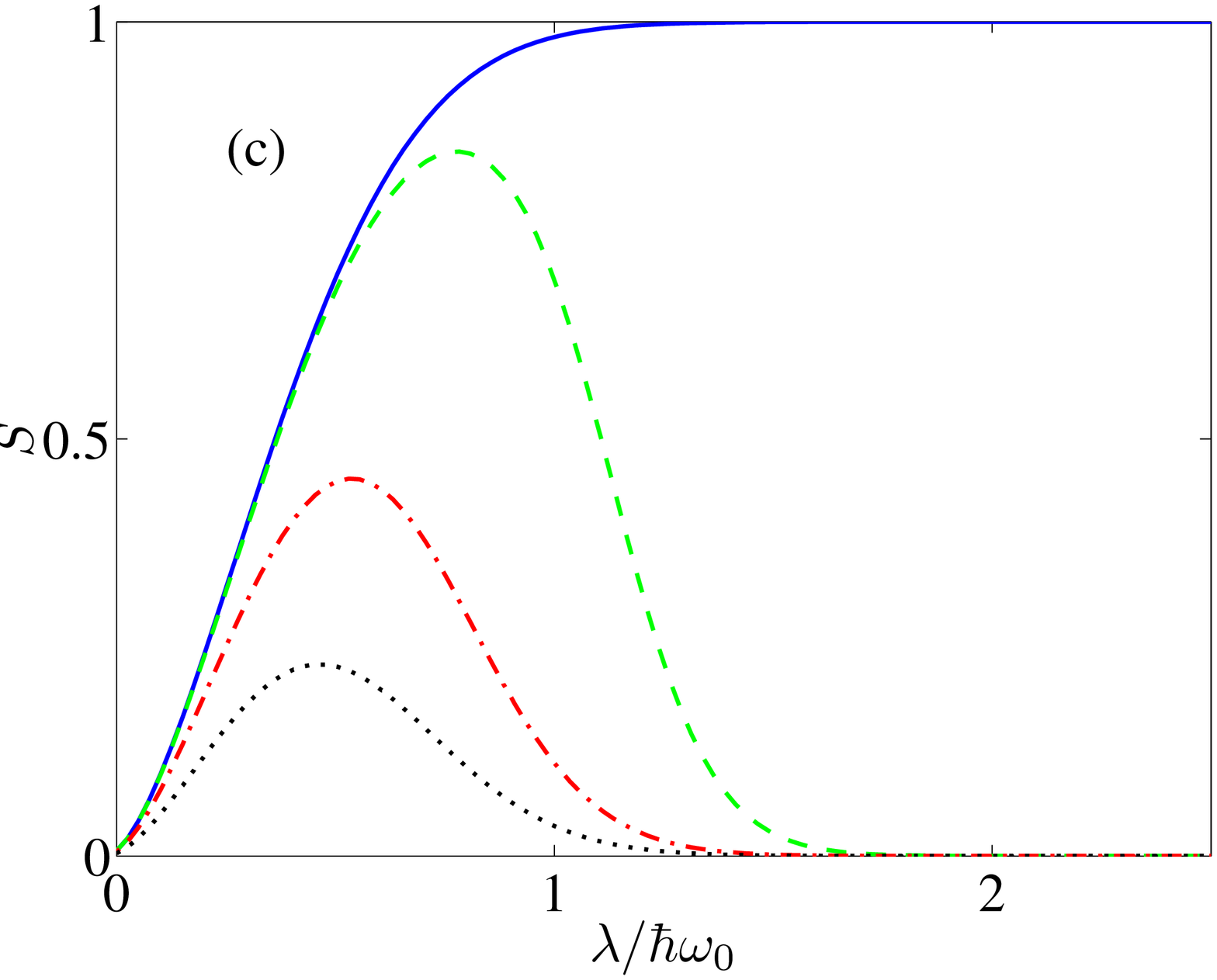}
\caption{(Color online) The qubit's entropy $S$ (which quantifies
the qubit-oscillator entanglement) in the ground state as a
function of $\lambda/(\hbar\omega_0)$. The ratio
$\hbar\omega_0/\Delta$ is $0.1$ in (a), 1 in (b) and 10 in (c),
and the different curves in each panel correspond to
$\epsilon/\Delta=0$ (blue, solid line), 0.1 (green, dashed line),
0.5 (red, dash-dotted line) and 1 (black, dotted line). For
$\epsilon=0$ the qubit-oscillator entanglement increases from zero
to one as $\lambda$ is increased, regardless of the relation
between $\hbar\omega_0$ and $E_{\rm q}$. However, the entanglement
drops rapidly (especially for large values of $\lambda$) as
$\epsilon$ is increased, i.e.~when the qubit is moved away from
the degeneracy point.} \label{Fig:QubitEntropy}
\end{figure}

In Fig.~\ref{Fig:QubitEntropy} we plot the qubit's ground-state
entropy as a function of $\lambda$. For $\epsilon=0$ the entropy
increases from zero to one as $\lambda$ increases from zero to
values much larger than all other parameters in the problem.
Demonstrating the fragility of the entangled states in the
large-$\lambda$ limit, Fig.~\ref{Fig:QubitEntropy} shows that the
entanglement drops rapidly (especially for large values of
$\lambda$) when $\epsilon$ is increased.

By comparing Figs.~\ref{Fig:Squeezing} and \ref{Fig:QubitEntropy},
we can see that the rise in the qubit-oscillator entanglement is
correlated with the reversal of the squeezing. One therefore goes
from a squeezed state in the oscillator to a qubit-oscillator
entangled state. We do not find any set of parameters where the
ground state contains an unentangled Schr\"odinger-cat state in
the oscillator.

The numerical results show that the case $\hbar\omega_0\ll E_{\rm
q}$ is most suited for the preparation of squeezed states, as can
be seen by comparing the maximum achievable squeezing in the
different parameter regimes. The opposite case ($\hbar\omega_0\gg
E_{\rm q}$) is most suited for the preparation of entangled
states, as seen from the extreme fragility of these states for the
case $\hbar\omega_0\ll E_{\rm q}$. In fact, all the nonclassical
properties of the ground state are suppressed as $\epsilon$ is
increased from zero to values larger than $\Delta$. We shall come
back to this point in Sec.~\ref{Sec:Decoherence}.

\begin{figure}[h]
\includegraphics[width=7.0cm]{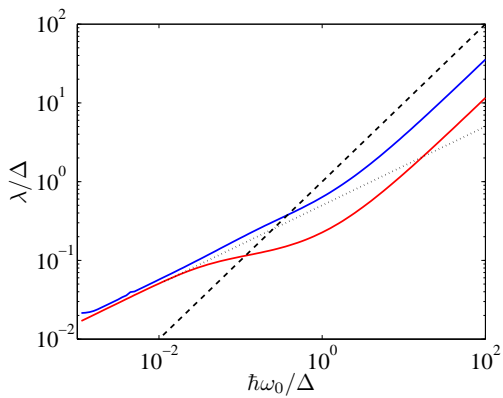}
\caption{(Color online) The value of $\lambda/\Delta$ at which the
qubit's ground-state entropy has the values 0.1 (red, lower solid
line) and 0.5 (blue, upper solid line) plotted as a function of
$\hbar\omega_0/\Delta$ (Note the logarithmic scale). Here we take
$\epsilon=0$. The straight lines are given by the formulae
$\lambda=\sqrt{\hbar\omega_0 E_{\rm q}}/2$ (dotted line) and
$\lambda=\hbar\omega_0$ (dashed line), which we have obtained in
Sec.~\ref{Sec:AnalyticalMethods}. For small values of
$\hbar\omega_0/\Delta$, the onset of entanglement occurs when
$\lambda=\sqrt{\hbar\omega_0 E_{\rm q}}/2$. For large  values of
$\hbar\omega_0/\Delta$, the onset of entanglement occurs when
$\lambda\sim\hbar\omega_0$, in agreement with the dependence
explained in Sec.~\ref{Sec:AnalyticalMethods}.}
\label{Fig:EntanglementOnset}
\end{figure}

In Fig.~\ref{Fig:EntanglementOnset} we examine the value of
$\lambda$ at which the qubit's ground-state entropy has the values
0.1 and 0.5. These curves serve as indicators for the onset of
qubit-oscillator entanglement, which is related to the instability
encountered in the semi-classical calculation. For a
high-frequency qubit, the sharp rise in entanglement occurs at
$\lambda=\sqrt{\hbar\omega_0 E_{\rm q}}/2$ which agrees with the
instability condition of
Secs.~\ref{Sec:AnalyticalMethodsAdiabaticQubit} and
\ref{Sec:AnalyticalMethodsSemiclassical}. For a high-frequency
oscillator, the entanglement rises to large values when
$\lambda\sim\hbar\omega_0$, in agreement with the analysis of
Sec.~\ref{Sec:AnalyticalMethodsAdiabaticOscillator}.

\section{Preparation and detection of nonclassical states through {\it in-situ}
parameter and state manipulation}
\label{Sec:PreparationAndDetectionInSitu}

We have seen in Sec.~\ref{Sec:NumericalCalculations} that
oscillator squeezed states and qubit-oscillator entangled states
can occur naturally as ground states of the strongly coupled
system. Schr\"odinger-cat states in the oscillator, i.e.~not
involving entanglement with the qubit, do not occur as ground
states of this system.

One method that has been proposed for the generation of oscillator
Schr\"odinger-cat states in the context of cavity QED
\cite{GerryWalls} can be considered here as well: One prepares a
qubit-oscillator entangled state of the form
\begin{equation}
\frac{1}{\sqrt{2}} \left( \ket{\alpha}\otimes\ket{q_1} +
\ket{-\alpha}\otimes\ket{q_2} \right),
\label{Eq:Qubit_oscillator_maximally_entangled_state}
\end{equation}
with $\ket{q_1}$ and $\ket{q_2}$ being any two orthogonal qubit
states and $\ket{\pm\alpha}$ being coherent states of the
oscillator with a large value of $|\alpha|$. One now measures the
qubit in the $(\ket{q_1}\pm\ket{q_2})/\sqrt{2}$ basis. Depending
on the outcome of the measurement, the state of the oscillator is
projected into one of the two states
\begin{equation}
\frac{1}{\sqrt{2}} \left( \ket{\alpha} \pm \ket{-\alpha} \right),
\end{equation}
each of which is a Schr\"odinger-cat state. Since the ground state
well above the critical point is approximately given by
\begin{equation}
\frac{1}{\sqrt{2}} \left(\ket{\alpha} \otimes\ket{\downarrow} +
\ket{-\alpha}\otimes\ket{\uparrow} \right),
\label{Eq:Qubit_oscillator_ground_state}
\end{equation}
with $\alpha=x_0$, the above procedure could also be implemented
in the system under consideration (We shall also give an
alternative procedure below). Hence all three types of
nonclassical states that we consider in this paper can be
generated in principle.

One important question that arises in the case of strong
qubit-oscillator coupling is whether it is possible to detect the
nonclassical states in spite of the always-present strong
coupling. The answer is yes, in principle. An important point to
note in this context is that, as shown in
Secs.~\ref{Sec:AnalyticalMethods} and
\ref{Sec:NumericalCalculations}, the energy eigenstates of the
system are approximately product states when the qubit is biased
far from the degeneracy point, i.e.~for large values of
$\epsilon$. One could therefore say that, for certain procedures,
the qubit and oscillator can be made to effectively decouple from
each other by biasing the qubit away from the degeneracy point.

We now discuss some possible procedures for the experimental
observation of the different nonclassical states. Since all three
types of nonclassical states occur in the case of a
high-frequency, adiabatically adjusting qubit, we focus on this
case. An experiment could start by preparing the ground state with
the qubit biased at the degeneracy point. Under suitable
conditions, this step could be achieved by biasing the qubit at
the degeneracy point and letting the system relax to its ground
state. The state would then be either a squeezed state or an
entangled state, depending on the coupling strength. One could
then move the qubit away from the degeneracy point for measurement
purposes. If the change is slow on the timescales of both the
qubit and the oscillator, the system will remain close to its
ground state, adiabatically following the bias-point shift (There
might be substantial excitation out of the ground state in the
double-well regime where the separation between the lowest energy
levels is small; however, this situation will come to an end when
the double-well potential transforms into a single-well
potential). If the system follows its instantaneous ground state
during the bias-point sweep, the nonclassical state will be lost.
The sweep therefore has to be fast at least compared to the period
of the oscillator. If the sweep is adiabatic with respect to the
qubit but fast with respect to the oscillator, the qubit will
follow the change adiabatically, while the oscillator will be
frozen in its initial state. In the case where the initial ground
state is a squeezed state, one would achieve the effective
qubit-oscillator decoupling while preserving the squeezed state
for the measurement step of the experiment. In the case where the
initial ground state is (approximately) the entangled state given
in Eq.~(\ref{Eq:Qubit_oscillator_ground_state}), the qubit will
end up in its ground state at the final bias point, and this state
will be independent of the state of the oscillator. As a result,
the oscillator is left in a Schr\"odinger-cat state. If the
bias-point sweep is fast on both the qubit and oscillator
timescales, both subsystems will be frozen in their initial state
during the sweep, such that one ends up with an entangled state at
the end of the sweep.

The state of the oscillator can be reconstructed using Wigner
tomography, which could be implemented following the experiment in
Ref.~\cite{Hofheinz}. In that experiment, the oscillator was put
into resonance with a qubit that was initialized in its ground
state, and the excitation probability of the qubit as a function
of time was determined by performing an ensemble of measurements.
By decomposing the signal into its Fourier components, it was
possible to extract the occupation probabilities of the different
photon-number states. When combined with the ability to shift the
oscillator's state (through the application of a classical driving
signal) before the measurement, full Wigner tomography becomes
possible. In the case of a low-frequency oscillator, the transfer
of excitations between the qubit and the oscillator can be induced
by driving the red or blue sideband, as was done in the experiment
of Ref.~\cite{Chiorescu}. Since, as explained above, the qubit is
effectively decoupled from the resonator when $\epsilon\gg\Delta$,
the exchange of excitations between the qubit and the oscillator
would not be efficient at the measurement bias point, which seems
to pose a dilemma for the proposed experiment. This difficulty can
be circumvented, however, by using a second, weakly coupled qubit
for measurement purposes.

Wigner tomography can be used to demonstrate squeezed and
Schr\"odinger-cat states in the oscillator. Qubit-oscillator
entangled states could be demonstrated by measuring the
correlation between the states of the qubit and oscillator.
Starting from the ground state, if the qubit is measured and found
to be in the state $\ket{\uparrow}$, the oscillator must be in the
state $\ket{-\alpha}$, with $\alpha=x_0$. If the qubit is found to
be in the state $\ket{\downarrow}$, the oscillator must be in the
state $\ket{\alpha}$. The observation of only these correlations,
however, is not sufficient in order to establish the presence of
quantum correlations. For that purpose one has to perform
measurements in more than one set of bases. The additional qubit
basis can be $(\ket{\uparrow}\pm\ket{\downarrow})/\sqrt{2}$: If
the qubit is found to be in the state
$(\ket{\uparrow}+\ket{\downarrow})/\sqrt{2}$, the oscillator must
be in the state $(\ket{\alpha}+\ket{-\alpha})/\sqrt{2}$, and a
similar relation holds for the minus signs. The two states
$(\ket{\alpha}\pm\ket{-\alpha})/\sqrt{2}$ can be distinguished
through the fact that the state with the plus sign contains only
even photon numbers while the state with the minus sign contains
only odd photon numbers.

Finally it should be noted that after the bias-point sweep, the
resulting state would not be a stationary state and would
therefore have a time dependence. This time dependence has to be
taken into account in the measurement sequence. Furthermore, when
the qubit is biased such that it is in one of the two
$\hat{\sigma}_z$ eigenstates, say $\ket{\uparrow}$, the effective
oscillator potential will be shifted from the point $x=0$, and one
must take into account this shift when analyzing the post-sweep
dynamics.

\section{decoherence}
\label{Sec:Decoherence}

We now turn to the question of how coupling to the environment
affects the prospects of preparing and observing nonclassical
states in the system under consideration. Following a standard
procedure \cite{CohenTannoudji,Ithier,DeLiberato}, we analyze the
effects of the environment by first determining the energy
eigenstates of the system in isolation and then analyzing the
relaxation and dephasing rates that govern the decoherence between
the different energy eigenstates. For our purposes it will be
sufficient to consider Markovian decoherence dynamics.

Since the preparation of nonclassical states above required
biasing the system at the point $\epsilon=0$, we focus on this
case. In Secs.~\ref{Sec:AnalyticalMethods} and
\ref{Sec:NumericalCalculations}, we found two types of low-lying
energy eigenstates, depending on the coupling strength. For small
values of $\lambda$, where the ground state involves a squeezed
state of the oscillator, the energy eigenstates are slightly
modified from the energy eigenstates in the absence of
qubit-oscillator coupling. For large values of $\lambda$, the
low-lying energy eigenstates are superpositions similar to that
given in Eq.~(\ref{Eq:Qubit_oscillator_ground_state}). As we shall
see below, there is a qualitative difference in how these two
types of states are affected by the environment.

The relaxation rate $\Gamma_{i \rightarrow j}$ and the dephasing
rate $\Gamma_{\varphi,ij}$ involving states $i$ and $j$ are given
by \cite{CohenTannoudji,Ithier}
\begin{eqnarray}
\Gamma_{i \rightarrow j} & = & \frac{\pi}{2} \; S \! \left(
\frac{E_i-E_j}{\hbar} \right) \times \left|\bra{i} \hat{A} \ket{j}
\right|^2 \nonumber
\\
\Gamma_{\varphi,ij} & = & \pi S(0) \times \left|\bra{i} \hat{A}
\ket{i} - \bra{j} \hat{A} \ket{j} \right|^2,
\label{Eq:Decoherence_rates}
\end{eqnarray}
where $S(\omega)$ is the environment's spectral density of the
relevant environment operator at frequency $\omega$, and $\hat{A}$
is the system operator through which the system couples to the
environment.

In order to go further with the analysis, we need to specify the
operator $\hat{A}$ that describes the coupling to the environment;
there is one such operator for each decoherence channel. It was
mentioned in the introduction that the availability of the tuning
parameter $\epsilon$ can be seen as an advantage of solid-state
qubits in comparison to natural atoms in cavity QED. On the other
hand, however, this property also opens an additional channel for
noise and the environment to couple to the system. The operator
associated with the parameter $\epsilon$ is $\hat{\sigma}_z$, and
the coupling through this operator is typically the most
detrimental for superconducting qubit circuits. We therefore start
by considering this decoherence channel.

For coupling through the operator $\hat{\sigma}_z$, the relaxation
and dephasing rates are proportional to the quantities
$\left|\bra{i} \hat{\sigma}_z \ket{j} \right|^2$ and
$\left|\bra{i} \hat{\sigma}_z \ket{i} - \bra{j} \hat{\sigma}_z
\ket{j} \right|^2$, respectively. For $\epsilon=0$ and small
$\lambda$ (and avoiding the resonant case), the energy eigenstates
are approximately product states with the qubit being in an
eigenstate of $\hat{\sigma}_x$, to which we refer as $\ket{\pm}$,
and the oscillator having a certain number of photons $n$. The
above quantities are then approximately given by the corresponding
values for the qubit states:
\begin{eqnarray}
\bra{n,+} \hat{\sigma}_z \ket{n',-} & = & \delta_{n,n'} \nonumber
\\
\bra{n,+} \hat{\sigma}_z \ket{n',+} & = & \bra{n,-} \hat{\sigma}_z
\ket{n',-} = 0.
\end{eqnarray}
These expressions suggest that the system relaxes with a rate
equal to that of the isolated qubit. The vanishing of the
dephasing rate in this approximation has the same origin as its
vanishing for an isolated qubit at the degeneracy point, namely
the fact that the energies are insensitive to variations in
$\epsilon$ to first order. This property points out an important
requirement for the above expressions to be valid: the
energy-level separation must be much larger than the transverse
fluctuations in the Hamiltonian. These fluctuations, which are
transverse at the degeneracy point, are responsible for dephasing
away from the degeneracy point. As a result, in order to obtain
the degeneracy-point protection from dephasing, the energy-level
separation must be large compared to the dephasing rate when the
latter is calculated away from the degeneracy point:
\begin{eqnarray}
\Gamma_{\varphi,ij} & = & \pi S(0) \times \left|\bra{n,\uparrow}
\hat{\sigma}_z \ket{n,\uparrow} - \bra{n',\downarrow}
\hat{\sigma}_z \ket{n',\downarrow} \right|^2 \nonumber
\\
& = & 4 \pi S(0).
\end{eqnarray}
This parameter is typically of the order of 100 MHz, which
corresponds to a dephasing time of 10 ns. At the degeneracy point,
an isolated qubit is protected from this noise because the qubit's
gap is typically larger than 1 GHz, and the qubit-environment
coupling is transverse to the qubit Hamiltonian. Similarly, mildly
squeezed states (whose energy-level separations are comparable to
those of the simple product states of the uncoupled system) should
be protected from dephasing caused by the weak, transverse
coupling to the environment. The situation is quite different for
large values of $\lambda$, where the low-lying states are highly
entangled states with a very small separation within an
energy-level pair. For example, when $\epsilon=0$, the lowest two
energy eigenstates are given by
\begin{equation}
\frac{1}{\sqrt{2}} \left(\ket{\alpha}\otimes\ket{\downarrow} \pm
\ket{-\alpha}\otimes\ket{\uparrow} \right),
\label{Eq:Qubit_oscillator_lowest_two_states}
\end{equation}
with $\alpha=x_0$. The energy separation between these states can
be obtained from the WKB calculation of
Sec.~\ref{Sec:AnalyticalMethodsAdiabaticQubit}. When $\epsilon$
exceeds this (possibly very small) energy separation, the energy
eigenstates are simply the product states $\ket{\alpha}
\otimes\ket{\downarrow}$ and $\ket{-\alpha}
\otimes\ket{\uparrow}$. In order for the entangled states to be
robust against fluctuations in $\epsilon$, their energy-level
separation (at $\epsilon=0$) must be much larger than the qubit's
dephasing rate of about 100 MHz. Since the qubit and oscillator
frequencies can be of the order of 1 GHz, one could obtain an
entangled ground state that is separated from the first excited
state by 100 MHz or more. For example, taking
$\lambda=\hbar\omega_0=\Delta=1$ GHz, we obtain a qubit
ground-state entropy of 0.85 and an energy separation of $138$
MHz. However, it would be highly desirable to reduce the qubit
decoherence rates, and in principle this should be possible in the
future using better materials and circuit designs.

Even though superconducting harmonic oscillators generally have
much higher quality factors than superconducting qubits, it is
instructive to briefly discuss the effect of oscillator
decoherence. The oscillator typically couples to its environment
through the same operator that describes its coupling to the
qubit. In the present problem, this operator is $\hat{x}$, or
equivalently $\hat{a}+\hat{a}^{\dagger}$. In a free oscillator,
the relevant matrix element for purposes of determining the
decoherence rates is given by
\begin{equation}
\bra{n} \left( \hat{a}+\hat{a}^{\dagger} \right) \ket{n'} =
\sqrt{n} \delta_{n-1,n'} + \sqrt{n+1} \delta_{n+1,n'}.
\end{equation}
Using the relations for the relaxation and dephasing rates in
Eq.~(\ref{Eq:Decoherence_rates}), one finds that at low
temperatures the effect of the environment is to cause decay to
the ground state through the loss of photons; photons are lost one
by one, with a rate that is proportional to the photon number
(Note that since $\bra{n} \left( \hat{a}+\hat{a}^{\dagger} \right)
\ket{n} = 0$, no pure dephasing occurs in this system). The
photon-loss process is described by the jump operator $\hat{a}$.
In the strongly coupled qubit-oscillator system with the
double-well effective potential, the effect of the environment
will be drastically different from the simple photon-loss
dynamics. Since the effective potential near each one of the local
minima has the same shape as the free-oscillator potential, the
energy eigenstates $\ket{n,\pm}$ will be entangled states where
the qubit is either in the state $\ket{\uparrow}$ or
$\ket{\downarrow}$, correlated with an oscillator wave function
that is given simply by the free-oscillator state $\ket{n}$
shifted to the left or right by a distance $x_0$. One therefore
finds the matrix elements
\begin{eqnarray}
\bra{n,\pm} \left( \hat{a}+\hat{a}^{\dagger} \right) \ket{n',\pm}
& = & \sqrt{n} \delta_{n-1,n'} + \sqrt{n+1} \delta_{n+1,n'}
\nonumber
\\
\bra{n,+} \left( \hat{a}+\hat{a}^{\dagger} \right) \ket{n',-} & =
& 0.
\end{eqnarray}
The fact that the matrix elements in the first line coincide with
those of the free oscillator implies that the relaxation rate will
be equal to the free-oscillator relaxation rate. Note that this
process is no longer described by the jump operator $\hat{a}$, but
rather by a properly shifted annihilation operator,
$\hat{a}\pm\sqrt{m\omega_0/(2\hbar)}x_0$, depending on whether one
is dealing with the left or right well in the effective
double-well potential. The relaxation process does not change the
state of the qubit, or any superposition involving the left and
right wells. Even though the relation
\begin{equation}
\bra{n,\pm} \left(\hat{a}+\hat{a}^{\dagger}\right) \ket{n,\pm} = 0
\end{equation}
would suggest that no dephasing should occur between the different
energy eigenstates, the alternative basis with localized states
$\{\ket{n,\uparrow},\ket{n,\downarrow}$ gives
\begin{equation}
\left| \bra{n,\uparrow} \left(\hat{a}+\hat{a}^{\dagger}\right)
\ket{n,\uparrow} - \bra{n,\downarrow}
\left(\hat{a}+\hat{a}^{\dagger}\right) \ket{n,\downarrow}
\right|^2 = \frac{8m\omega_0}{\hbar} x_0^2.
\end{equation}
This result implies that the coupling to the environment will
cause dephasing in any quantum superposition involving the two
wells, with a tendency to localize the wave function in one of the
wells. The rate of this process will be proportional to the
product of the environment's power spectrum at zero frequency and
the combination of matrix elements given above. This last quantity
is proportional to $x_0^2$, and it grows indefinitely with
increasing coupling strength. If we assume that the power spectrum
at zero frequency is comparable to that at the oscillator
frequency, the dephasing rate can be much larger than the decay
rate of the free oscillator because of the largeness of the
quantity $\left| \bra{n,\uparrow}
\left(\hat{a}+\hat{a}^{\dagger}\right) \ket{n,\uparrow} -
\bra{n,\downarrow} \left(\hat{a}+\hat{a}^{\dagger}\right)
\ket{n,\downarrow} \right|^2$. For example taking the oscillator's
decay rate to be 0.1-1 MHz and considering states where $\hat{a}$
and $\hat{a}^{\dagger}$ are of typical size $\sqrt{5}$ (i.e.~about
five virtual photons in the ground state), we find that the
dephasing rate can be of the order of 10-100 MHz, which is
comparable to the $\hat{\sigma}_z$-mediated dephasing rate.

We finally discuss the question of temperature. We have considered
the possibility of preparing nonclassical states by letting the
system cool down to its ground state. One must therefore make sure
that the energy-level separation between the ground state and the
first excited state is larger than the ambient temperature. In
superconducting circuits, the temperature is typically around 20
mK, which can be converted to roughly 1 GHz in frequency units.
The ground state must be separated from the excited states by at
least that amount in order to achieve high-fidelity preparation of
the ground state. Typical qubit and oscillator frequencies are in
the few-gigahertz range, not much higher than typical
temperatures. Squeezed ground states in the oscillator should be
separated from the excited states by an energy comparable to the
one present in the uncoupled system, implying that the preparation
of these states should be possible. The entangled ground states
that occur for strong coupling, however, are separated from the
first excited states by energy gaps that decrease rapidly with
increasing qubit-oscillator coupling strength. If this energy gap
becomes smaller than the 1 GHz temperature level, one would not be
able to prepare the entangled ground state simply by letting the
system cool down to such a state. However, one could let the
system cool down to its ground state away from the degeneracy
point and then adiabatically shift the bias point to one with an
entangled ground state. Provided that the thermalization rate is
sufficiently low, it is not necessary to have a degeneracy-point
energy-level separation that is larger than the temperature. The 1
GHz temperature level should therefore not be seen as a
fundamental obstacle to the preparation of entangled ground
states.

\section{Conclusion}
\label{Sec:Conclusion}

We have analyzed the properties of a strongly coupled
qubit-oscillator system, focusing on the potential of this system
for the preparation of nonclassical states. These states include
squeezed states and Schr\"odinger-cat states of the oscillator, as
well as qubit-oscillator entangled states.

We have compared the predictions of four different analytical
approaches: the weak-coupling approximation, the
adiabatically-adjusting-oscillator approximation, the
adiabatically-adjusting-qubit approximation and the semiclassical
calculation. Each one of these four approaches is well suited for
analyzing the behaviour of the system under a certain set of
assumptions. Thus the combination of the results provides a rather
thorough understanding of the qubit-oscillator system in the
regime of ultrastrong coupling. We have also presented results of
numerical calculations that reinforce the results of the
analytical derivations. These results demonstrate the nonclassical
properties of the energy eigenstates, and especially the ground
state, of the system.

We have discussed various possible experimental procedures for the
preparation and observation of nonclassical states. All three
types of nonclassical states that we discuss in this paper can be
prepared and detected in principle.

We have also analyzed the effect of coupling to the environment on
the system. We have shown that the decoherence dynamics of the
coupled qubit-oscillator system can be qualitatively different
from the decoherence dynamics of the qubit or oscillator in
isolation. We have shown that nonclassical states, particularly
highly-entangled states, are highly susceptible to changes or
fluctuations in the bias parameters. These results lead to the
conclusion that high degrees of control and low noise levels will
be required for the preparation of robust nonclassical states in
the ultrastrong-coupling regime.

We would like to thank P. Forn-Diaz, J. R. Johansson, N. Lambert
and S. Shevchenko for useful discussions. This work was supported
in part by the National Security Agency (NSA), the Army Research
Office (ARO), the Laboratory for Physical Sciences (LPS) and the
National Science Foundation (NSF) grant No.~0726909.

{\it Note added in proof}: Recently ultrastrong coupling between a
superconducting flux qubit and a coplanar waveguide resonator has
been demonstrated \cite{Niemczyk}.

\begin{center}
{\bf Appendix A: Oscillator's effective kinetic energy in the case of a high-frequency qubit}
\end{center}

In this appendix we briefly discuss the modification to the
oscillator's kinetic energy in the case of a high-frequency qubit
(see Sec.~\ref{Sec:AnalyticalMethodsAdiabaticQubit}). One could
perform the calculation by considering only the kinetic-energy
term, without considering the specific form of the trapping
potential, i.e.~by using the completely delocalized momentum
eigenstates as a basis for the calculation. Such a calculation,
however, turns out not to lead to simple, transparent results. We
therefore consider the corrections that one would need to add to
the kinetic-energy term in the effective oscillator Hamiltonian
starting from the eigenstates of the free-oscillator Hamiltonian.

The eigenstates of the Hamiltonian $\hat{H}_{\rm ho}$ can be
expressed in the position basis as
\begin{equation}
\ket{k} = \int dx \psi_k(x) \ket{x}.
\end{equation}
Taking into account the high-frequency, adiabatically adjusting
qubit, and for definiteness taking the case of the qubit's ground
state, the above eigenstates are modified as follows:
\begin{eqnarray}
\ket{\tilde{k}} & = & \int dx \psi_k(x) \ket{x} \otimes \ket{g(x)}
\nonumber
\\
& = & \int dx \psi_k(x) \sum_{k'} \psi_{k'}^*(x) \ket{k'} \otimes
\ket{g(x)},
\end{eqnarray}
where $\ket{g(x)}$ has the same meaning as $\ket{g_x}$ from
Sec.~\ref{Sec:AnalyticalMethodsAdiabaticQubit}. One can now obtain
the correction to the kinetic-energy term as:
\begin{eqnarray}
& & \!\!\!\!\! \bra{\tilde{k}} \frac{\hat{p}^2}{2m}
\ket{\tilde{l}} - \bra{k} \frac{\hat{p}^2}{2m} \ket{l} =
\int\!\!\int dx_1 dx_2 \sum_{k'l'} \psi_k^*(x_1) \psi_{k'}(x_1)
\nonumber
\\
& & \hspace{0.8cm} \psi_{l'}^*(x_2) \psi_l(x_2)
\bra{k'}\frac{\hat{p}^2}{2m} \ket{l'} \left(
\bra{g(x_1)}g(x_2)\rangle - 1 \right).
\end{eqnarray}
The factor between parentheses represents the relative correction
to the kinetic-energy term in the Hamiltonian. This factor can be
estimated as:
\begin{eqnarray}
\bra{g(x_1)}g(x_2)\rangle - 1 & = & \cos \left(
\frac{\varphi(x_1)-\varphi(x_2)}{2} \right) - 1 \nonumber
\\
& \approx & - \frac{\left[ \varphi(x_1)-\varphi(x_2) \right]^2}{8}
\nonumber
\\
& \sim & \left( \frac{g (x_1-x_2)}{E_{\rm q}} \right)^2 \nonumber
\\
& \sim & \frac{\hbar g^2}{m\omega_0 E_{\rm q}^2}.
\end{eqnarray}
In deriving this expression we have taken the case
$\epsilon/\Delta\ll 1$ (for which the relative correction is
maximum) and taken $x$ to be of the order of the characteristic
oscillator length. By comparing the above expression to the
relative correction in the potential-energy term found in
Sec.~\ref{Sec:AnalyticalMethodsAdiabaticQubit}, i.e. $g^2/(m
\omega_0^2 E_{\rm q})$, one can see that the kinetic-term
correction is negligible when $\hbar\omega_0\ll E_{\rm q}$.


\begin{thebibliography}{99}

\bibitem{JaynesCummings} E. T. Jaynes and F. W. Cummings, Proc. IEEE
{\bf 51}, 89 (1963).

\bibitem{GerryWalls} See e.g. C. C. Gerry and P. L. Knight, {\it
Introductory Quantum Optics} (Cambridge University Press, 2005);
D. F. Walls and G. J. Milburn, {\it Quantum Optics} (Springer,
Berlin, 1994); M. O. Scully and M. S. Zubairy, {\it Quantum
Optics} (Cambridge University Press, 1997).

\bibitem{Chiorescu} I. Chiorescu, P. Bertet, K. Semba, Y. Nakamura,
C. J. P. M. Harmans, J. E. Mooij, Nature {\bf 431}, 159 (2004).

\bibitem{Wallraff} A. Wallraff, D. I. Schuster, A. Blais, L. Frunzio,
R.-S. Huang, J. Majer, S. Kumar, S. M. Girvin, R. J. Schoelkopf,
Nature {\bf 431}, 162 (2004).

\bibitem{You} For recent reviews on superconducting qubit circuits, see
e.g. J. Q. You and F. Nori, Phys. Today {\bf 58} (11), 42 (2005);
G. Wendin and V. Shumeiko, in {\it Handbook of Theoretical and
Computational Nanotechnology}, ed. M. Rieth and W. Schommers (ASP,
Los Angeles, 2006); R. J. Schoelkopf and S. M. Girvin, Nature,
{\bf 451}, 664 (2008); J. Clarke and F. K. Wilhelm, Nature {\bf
453}, 1031 (2008).

\bibitem{Hennessy} K. Hennessy, A. Badolato, M. Winger, D. Gerace,
M. Atat\"ure, S. Gulde, S. F\"alt, E. L. Hu, and A. Imamoglu,
Nature (London) {\bf 445}, 896 (2007).

\bibitem{Leturcq} R. Leturcq, C. Stampfer, K. Inderbitzin, L. Durrer,
C. Hierold, E. Mariani, M. G. Schultz, F. von Oppen, and K.
Ensslin, Nat. Phys. {\bf 5}, 327 (2009); M. D. LaHaye, J. Suh, P.
M. Echternach, K. C. Schwab, and M. L. Roukes, Nature {\bf 459},
960 (2009); G. A. Steele, A. K. H\"uttel, B. Witkamp, M. Poot, H.
B. Meerwaldt, L. P. Kouwenhoven, and H. S. J. van der Zant,
Science {\bf 325}, 1103 (2009).

\bibitem{Armour} A. D. Armour, M. P. Blencowe, and K. C. Schwab, Phys. Rev.
Lett. {\bf 88}, 148301 (2002).

\bibitem{Holstein} T. Holstein, Ann. Phys. (N.Y.) {\bf 8}, 325 (1959).

\bibitem{Graham} See e.g. R. Graham and M. H\"ohnerbach, Z. Phys {\bf 57}, 233
(1984); L. M\"uller, J. Stolze, H. Leschke, and P. Nagel, Phys.
Rev. A {\bf 44}, 1022 (1991).

\bibitem{Kimble} H. J. Kimble, Phys. Scr. {\bf T76}, 127 (1998); J. M.
Raimond, M. Brune, and S. and Haroche, Rev. Mod. Phys. {\bf 73},
565 (2001); S. Haroche and J. M. Raimond, {\it Exploring the
Quantum: Atoms, Cavities, and Photons} (Oxford University Press,
2006).

\bibitem{Devoret} M. H. Devoret. S. Girvin and R. Schoelkopf, Ann. Phys.
{\bf 16}, 767 (2007).

\bibitem{Hines} A. P. Hines, C. M. Dawson, R. H. McKenzie, and G. J.
Milburn, Phys. Rev. A {\bf 70}, 022303 (2004).

\bibitem{Irish} E. K. Irish, J. Gea-Banacloche, I. Martin, and K. C.
Schwab, Phys. Rev. B {\bf 72}, 195410 (2005); E. K. Irish, Phys.
Rev. Lett. {\bf 99}, 173601 (2007).

\bibitem{Larson} J. Larson, Phys. Scr. {\bf 76}, 146 (2007); Phys. Rev. A
{\bf 78}, 033833 (2008); arXiv:0908.1717.

\bibitem{Zueco} D. Zueco, G. M. Reuther, S. Kohler, and P.
H\"anggi, Phys. Rev. A {\bf 80}, 033846 (2009).

\bibitem{Meaney} C. P. Meaney, T. Duty, R. H. McKenzie, and G. J. Milburn,
arXiv:0903.2681.

\bibitem{Lizuain} I. Lizuain, J. Casanova, J. J. Garcia-Ripoll, J. G. Muga,
and E. Solano, arXiv:0912.3485.

\bibitem{Oliver} See e.g. W. D. Oliver, Y. Yu, J. C. Lee, K. K. Berggren,
L. S. Levitov, T. P. Orlando, Science {\bf 310}, 1653 (2005); D.
M. Berns, W. D. Oliver, S. O. Valenzuela, A. V. Shytov, K. K.
Berggren, L. S. Levitov, and T. P. Orlando, Phys. Rev. Lett. {\bf
97}, 150502 (2006).

\bibitem{Paauw} See e.g. F. G. Paauw, A. Fedorov, C. J. P. M. Harmans, and
J. E. Mooij, Phys. Rev. Lett. {\bf 102}, 090501 (2009).

\bibitem{Hofheinz} M. Hofheinz, E. M. Weig, M. Ansmann, R. C. Bialczak, E.
Lucero, M. Neeley, A. D. O'Connell, H. Wang, J. M. Martinis, and
A. N. Cleland, Nature {\bf 454}, 310 (2008); M. Hofheinz, H. Wang,
M. Ansmann, R. C. Bialczak, E. Lucero, M. Neeley, A. D. O'Connell,
D. Sank, J. Wenner, J. M. Martinis and A. N. Cleland, Nature {\bf
459}, 546 (2009).

\bibitem{Houck} See also A. A. Houck, D. I. Schuster, J. M. Gambetta, J. A.
Schreier, B. R. Johnson, J. M. Chow, L. Frunzio, J. Majer, M. H.
Devoret, S. M. Girvin, and R. J. Schoelkopf, Nature {\bf 449}, 328
(2007); M. A. Sillanp\"a\"a, J. I. Park, and R.W. Simmonds, Nature
{\bf 449}, 438 (2007).

\bibitem{WeakCouplingProposals} For recent proposals for the generation of
squeezed and Schr\"odinger-cat states in superconducting
resonators, see e.g.~Y.-X. Liu, L.-F. Wei, and F. Nori, Europhys.
Lett. {\bf 67}, 941 (2004); Phys. Rev. A {\bf 71}, 063820 (2005);
M. Mariantoni, F. Deppe, A. Marx, R. Gross, F. K. Wilhelm, and E.
Solano, Phys. Rev. B {\bf 78}, 104508 (2008); A. M. Zagoskin, E.
Il'ichev, M. W. McCutcheon, J. F. Young, and F. Nori, Phys. Rev.
Lett. {\bf 101}, 253602 (2008); M. Abdel-Aty, Opt. Commun. {\bf
282}, 4556 (2009); J. R. Johansson, G. Johansson, C. M. Wilson,
and F. Nori, Phys. Rev. Lett. {\bf 103}, 147003 (2009).

\bibitem{Bourassa} J. Bourassa, J. M. Gambetta, A. A. Abdumalikov, O. Astafiev,
Y. Nakamura, and A. Blais, Phys. Rev. A {\bf 80}, 032109 (2009).

\bibitem{Peropadre} B. Peropadre, P. Forn-Diaz, E. Solano, and J. J.
Garcia-Ripoll, arXiv:0912.3456.

\bibitem{Shevchenko} See e.g. S. N. Shevchenko, S. Ashhab, and F. Nori, Phys.
Rep. (in press).

\bibitem{Sillanpaa} M. A. Sillanp\"a\"a, T. Lehtinen, A. Paila, Yu. Makhlin,
L. Roschier, and P. J. Hakonen, Phys. Rev. Lett. {\bf 95}, 206806
(2005); T. Duty, G. Johansson, K. Bladh, D. Gunnarsson, C. Wilson,
and P. Delsing, Phys. Rev. Lett. {\bf 95}, 206807 (2005); G.
Johansson, L. Tornberg, and C. M.Wilson, Phys. Rev. B {\bf 74},
100504(R) (2006); V. I. Shnyrkov, Th. Wagner, D. Born, S. N.
Shevchenko, W. Krech, A. N. Omelyanchouk, E. Il'ichev, and H.-G.
Meyer, Phys. Rev. B {\bf 73}, 024506 (2006); S. N. Shevchenko, S.
H. W. van der Ploeg, M. Grajcar, E. Il'ichev, A. N. Omelyanchouk,
H.-G. Meyer, Phys. Rev. B {\bf 78}, 174527 (2008).

\bibitem{CohenTannoudji} See e.g. C. Cohen-Tannoudji, J. Dupont-Roc, and G.
Grynberg, {\it Atom-Photon Interactions} (Wiley, New York, 1992).

\bibitem{Hausinger} An explanation of Van Vleck
perturbation theory can be found in J. Hausinger and M. Grifoni,
New J. Phys. {\bf 10}, 115015 (2008); An application to
superconducting-qubit circuits with high-frequency couplers is
given in S. Ashhab, A. O. Niskanen, K. Harrabi, Y. Nakamura, T.
Picot, P. C. de Groot, C. J. P. M. Harmans, J. E. Mooij, and F.
Nori, Phys. Rev. B {\bf 77}, 014510 (2008).

\bibitem{Dicke} R. H. Dicke, Phys. Rev. {\bf 93}, 99 (1954).

\bibitem{Hepp} K. Hepp and E. H. Lieb, Ann. Phys. {\bf 76}, 360
(1973); Y. K. Wang and F. T. Hioe, Phys. Rev. A {\bf 7}, 831
(1973); See also N. Lambert, C. Emary, and T. Brandes, Phys. Rev.
Lett. {\bf 92}, 073602 (2004); P. Nataf and C. Ciuti, Phys. Rev.
Lett. {\bf 104}, 023601 (2010).

\bibitem{Ithier} G. Ithier, E. Collin, P. Joyez, P. J. Meeson, D.
Vion, D. Esteve, F. Chiarello, A. Shnirman, Y. Makhlin, J.
Schriefl, and G. Sch\"on, Phys. Rev. B {\bf 72}, 134519 (2005).

\bibitem{DeLiberato} For a recent study on a related problem, see e.g. S.
De Liberato, D. Gerace, I. Carusotto, and C. Ciuti, Phys. Rev. A
{\bf 80}, 053810 (2009).

\bibitem{Niemczyk} T. Niemczyk, F. Deppe, H. Huebl, E. P. Menzel, F.
Hocke, M. J. Schwarz, J. J. Garcia-Ripoll, D. Zueco, T. H\"ummer,
E. Solano, A. Marx, and R. Gross, arXiv:1003.2376v1.


\end{thebibliography}
\end{document}